\documentclass[12pt,tightenlines,superscriptaddress,nofootinbib,colorlinks = true, linkcolor = blue, urlcolor = blue, citecolor = blue, anchorcolor = blue]{revtex4-2}
\usepackage{graphicx}
\usepackage[margin=1.25in]{geometry}
\usepackage[usenames,dvipsnames]{color}
\usepackage{url}
\usepackage[colorlinks = true,
            linkcolor = blue,
            urlcolor  = blue,
            citecolor = blue,
            anchorcolor = blue]{hyperref}

\usepackage{caption}
\usepackage{subcaption}
\usepackage{amssymb}
\usepackage{slashed}

\def\beq{\begin{equation}}
\def\eeq#1{\label{#1}\end{equation}}
\def\eeqn{\end{equation}}

\newenvironment{Eqnarray}%
   {\arraycolsep 0.14em\begin{eqnarray}}{\end{eqnarray}}
\def\beqa{\begin{Eqnarray}}
\def\eeqa#1{\label{#1}\end{Eqnarray}}
\def\eeqan{\end{Eqnarray}}

\let\bar=\overbar

\def\lsim{\mathrel{\raise.3ex\hbox{$<$\kern-.75em\lower1ex\hbox{$\sim$}}}}
\def\gsim{\mathrel{\raise.3ex\hbox{$>$\kern-.75em\lower1ex\hbox{$\sim$}}}}

\def\del{\partial}
\def\Dslash{\not{\hbox{\kern-4pt $D$}}}
\def\dslash{\not{\hbox{\kern-2pt $\del$}}}
\def\pslash{\not{\hbox{\kern-2pt $p$}}}
\def\ETmiss{\not{\hbox{\kern-4pt $E$}}_T}

\def\Dlr{\mathrel{\raise1.5ex\hbox{$\leftrightarrow$\kern-1em\lower1.5ex\hbox{$D$}}}}

\def\MSB{{\bar{M \kern -2pt S}}}
\def\msb{{\bar{\scriptsize M \kern -1pt S}}}

\def\drb{{\bar{\scriptsize D \kern -1pt R}}}

\def\eps{\epsilon}

\def\MeV{{\rm MeV}}

\newcommand\snowmass{\begin{center}\rule[-0.2in]{\hsize}{0.01in}\\\rule{\hsize}{0.01in}\\
\vskip 0.1in Submitted to the  Proceedings of the US Community Study\\ 
on the Future of Particle Physics (Snowmass 2021)\\ 
\rule{\hsize}{0.01in}\\\rule[+0.2in]{\hsize}{0.01in} \end{center}}

\usepackage{framed}

\newcommand{\temporarypagebreak}{{\newpage}}
\usepackage{xcolor}

\newcommand{\affilFNAL}{\affiliation{Fermi National Accelerator Laboratory, Batavia, IL 60510, USA}}

\makeatletter
\def\l@subsection#1#2{}
\def\l@subsubsection#1#2{}
\makeatother

\begin{document}


{\hfill FERMILAB-PUB-22-497-T}
\medskip


\title{A Snowmass Whitepaper: Dark Matter Production at Intensity-Frontier Experiments}

\author{Contacts and lead editors: G. Krnjaic}
\affilFNAL

\affiliation{Department of Astronomy and Astrophysics, University of Chicago,, Chicago, IL 60637}
\affiliation{Kavli Institute for Cosmological Physics, University of Chicago, Chicago, IL 60637}

\author{N. Toro}
\affiliation{SLAC National Accelerator Laboratory, Menlo Park, CA 94025, USA}

\author{\hfill \\ \hfill \\ Plot and Sub-Section Editors: A. Berlin}
\affilFNAL
\author{B. Batell}
\affiliation{Department of Physics and Astronomy, University of Pittsburgh, 3941 O'Hara St, Pittsburgh, PA 15260, USA}
\author{N. Blinov}
\affiliation{University of Victoria, Victoria, BC V8P 5C2, Canada}
\author{L. Darm\'e}
\affiliation{Institut de Physique des 2 Infinis de Lyon (IP2I), UMR5822, CNRS/IN2P3, 69622 Villeurbanne Cedex, France}
\author{P. DeNiverville}
\affiliation{Los Alamos National Laboratory, Los Alamos, NM 87545, USA}
\author{P. Harris}
\affiliation{Massachusetts Institute of Technology, Cambridge, MA 02139}
\author{C. Hearty}
\affiliation{Department of Physics and Astronomy, University of British Columbia (UBC), Vancouver, British
Columbia, V6T 1Z1 Canada}
\affiliation{Institute of Particle Physics (Canada), Victoria, British Columbia V8W 2Y2, Canada}
\author{M. Hostert}
\affiliation{School of Physics and Astronomy, University of Minnesota, Minneapolis, MN 55455, USA}
\affiliation{William I. Fine Theoretical Physics Institute, School of Physics and Astronomy, University of Minnesota, Minneapolis, MN 55455, USA}
\affiliation{Perimeter Institute for Theoretical Physics, Waterloo, ON N2J 2W9, Canada}
\author{K.J. Kelly}
\affiliation{European Organization for Nuclear Research (CERN), 1211 Geneva 23, Switzerland}
\author{D. McKeen}
\affiliation{TRIUMF, 4004 Wesbrook Mall, Vancouver, BC V6T 2A3, Canada}
\author{S. Trojanowski}
\affiliation{Astrocent, Nicolaus Copernicus Astronomical Center Polish Academy of Sciences, ul. Rektorska 4, 00-614, Warsaw, Poland}
\affiliation{National Centre for Nuclear Research, Pasteura 7, 02-093 Warsaw, Poland}
\author{Y.-D. Tsai}
\affiliation{Department of Physics and Astronomy, University of California, Irvine, CA 92697-4575, USA}

\begin{abstract}
\noindent 
Dark matter particles can be observably produced at intensity-frontier experiments, and opportunities in the next decade will explore important parameter space motivated by thermal DM models, the dark sector paradigm, and anomalies in data.  This whitepaper describes the motivations, detection strategies, prospects and challenges for such searches, as well as synergies and complementarity both within RF6 and across HEP.
 
\snowmass
\end{abstract}

\maketitle
\tableofcontents
\newpage

\newpage


\section*{Executive Summary}
\subsection*{Context and Motivation}
The existence of dark matter (DM) is Nature's sharpest evidence that the Standard Model (SM) of particle physics is incomplete. While astrophysical evidence for DM has mounted steadily over the past 8 decades, with increasingly precise measurements confirming the effects of DM from galactic and cluster scales to the primordial early Universe, its particle nature remains elusive.  Identifying the fundamental constituents of dark matter, how these came to dominate the matter density of the Universe, and how they connect to the well-understood physics of ordinary matter are arguably the greatest questions in fundamental physics today.

The space of possible DM masses and properties is vast: the range of viable masses for individual DM constituents spans roughly 50 orders of magnitude; to date its only observed interaction is through gravity; particle properties of DM have not been measured but bulk properties, including its cosmological mass density, inform and motivate models for the DM constituents.   
Indeed, the observed DM density has long served as a goalpost for understanding plausible models of DM, and a hint that suggests DM has microscopic interactions with ordinary matter that are stronger than gravity.  Early thermal equilibrium of DM and familiar matter, followed by freeze-out of the DM as the Universe cools, offers one simple explanation for the origin of its observed abundance.  This freeze-out mechanism is exemplified by the Weakly Interacting Massive Particle (WIMP) paradigm, which has long been the focus of terrestrial searches for dark matter. 

The last decade has seen a tremendous growth of theoretical and experimental interest in DM whose constituents are comparable in mass to electrons or protons, so-called ``light DM''.  This framework simply generalizes the WIMP paradigm to lower masses.  Light DM maintains the simplicity of thermal freeze-out as an origin for DM, as well as the close structural resemblance of the DM sector to the SM, yet poses different experimental challenges and opportunites.  Models of light DM rely for freeze-out on light force-carriers with parametrically weak SM couplings.  As a typical example, a new $U(1)$ gauge boson (``dark photon'') can mix with the SM photon at the $\sim 10^{-3}$ to $10^{-6}$ level due to radiative effects --- a degree of mixing compatible with thermal freeze-out for MeV to GeV DM.  These interactions are too weak to be detectable in high-$p_T$ DM searches at high-energy colliders, and the lighter DM particles carry too little kinetic energy to be seen in traditional direct detection.  

\subsection*{Accelerator Production of Light Dark Matter}
In response to these challenges, laboratory production of Light DM by intensity-frontier experiments --- including dedicated fixed-target experiments, small forward detectors, and flavor factories --- has emerged as an essential strategy for exploring light DM.  These experiments are optimized for intensity, instrumentation precision, and/or background rejection rather than energy reach.  

\begin{figure}[!ht]
    \begin{center}
    \includegraphics[width=0.9\hsize]{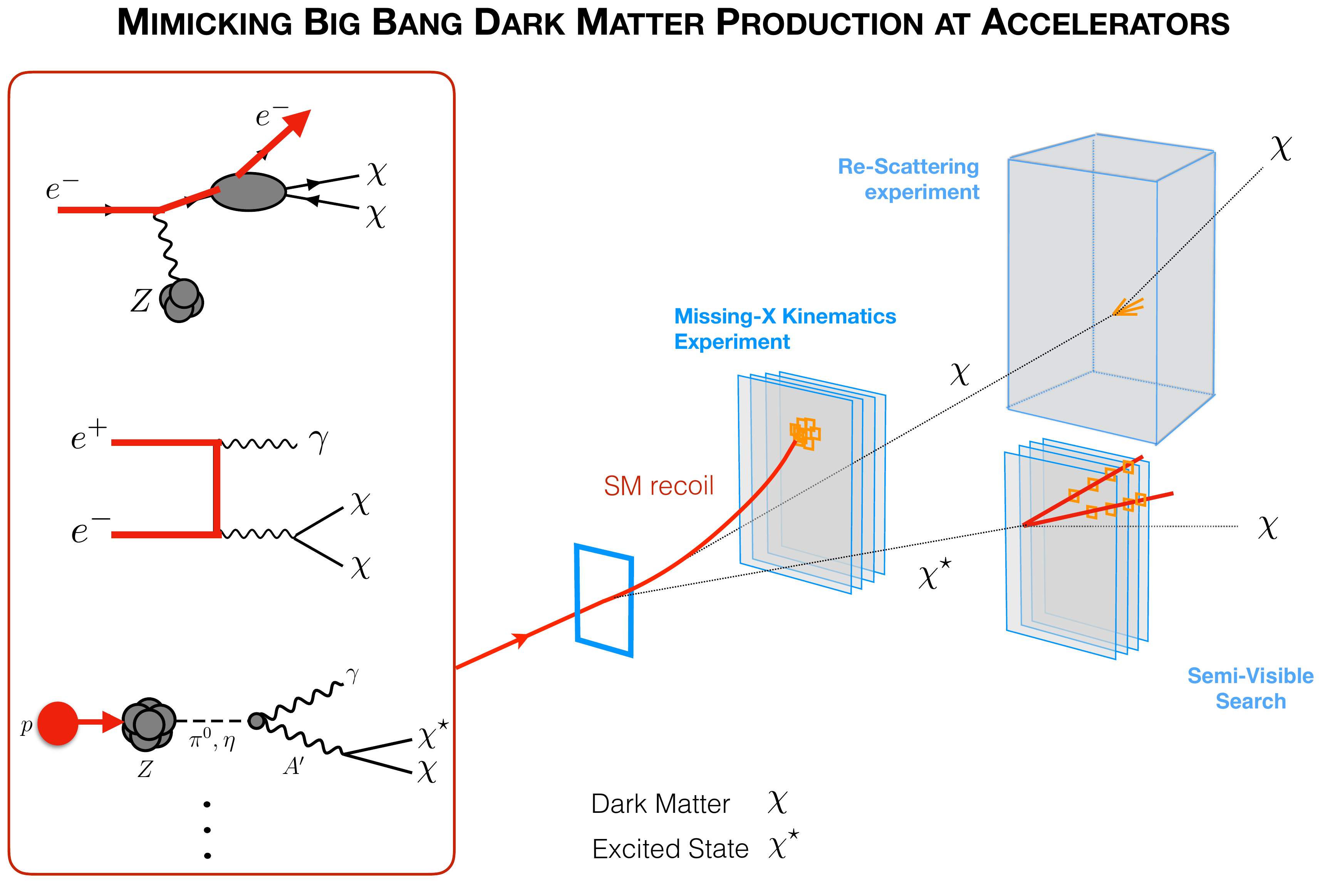}
    \end{center}
    \caption{Illustration of representative DM production mechanisms (left) and (right) the concepts for detecting DM production via, clockwise from left, missing X, re-scattering, and semi-visible detection strategies. 
    \label{fig:strategyCartoon}}
\end{figure}

Accelerator-based searches for dark matter exploit several different production mechanisms, including bremsstrahlung-like DM production off beam leptons or protons, meson decays that include DM in the final state, $e^+e^-$ annihilation, and Drell-Yan production.  The search strategies can be grouped into three broad categories:  \emph{Missing energy, momementum, or mass} searches use the kinematics of visible particles recoiling from a DM production event, together with vetoes on SM reaction products, to identify DM production events.  \emph{Re-scattering} experiments search for DM and/or millicharged particles through their subsequent scattering in a detector forward of a fixed-target beam-dump or collider interaction point.  \emph{Semi-visible searches} leverage the possibility of metastable resonances in the dark-sector, which can be motivated by specific models of DM cosmology and in many cases decay into a combination of DM and visible SM particles.  These strategies are summarized visually in Fig. \ref{fig:strategyCartoon} and the program has been reviewed in  Refs. \cite{Essig:2013lka,Battaglieri:2017aum}.   

\begin{figure}[t!]
    \begin{center}
        \includegraphics[width=0.9\hsize]{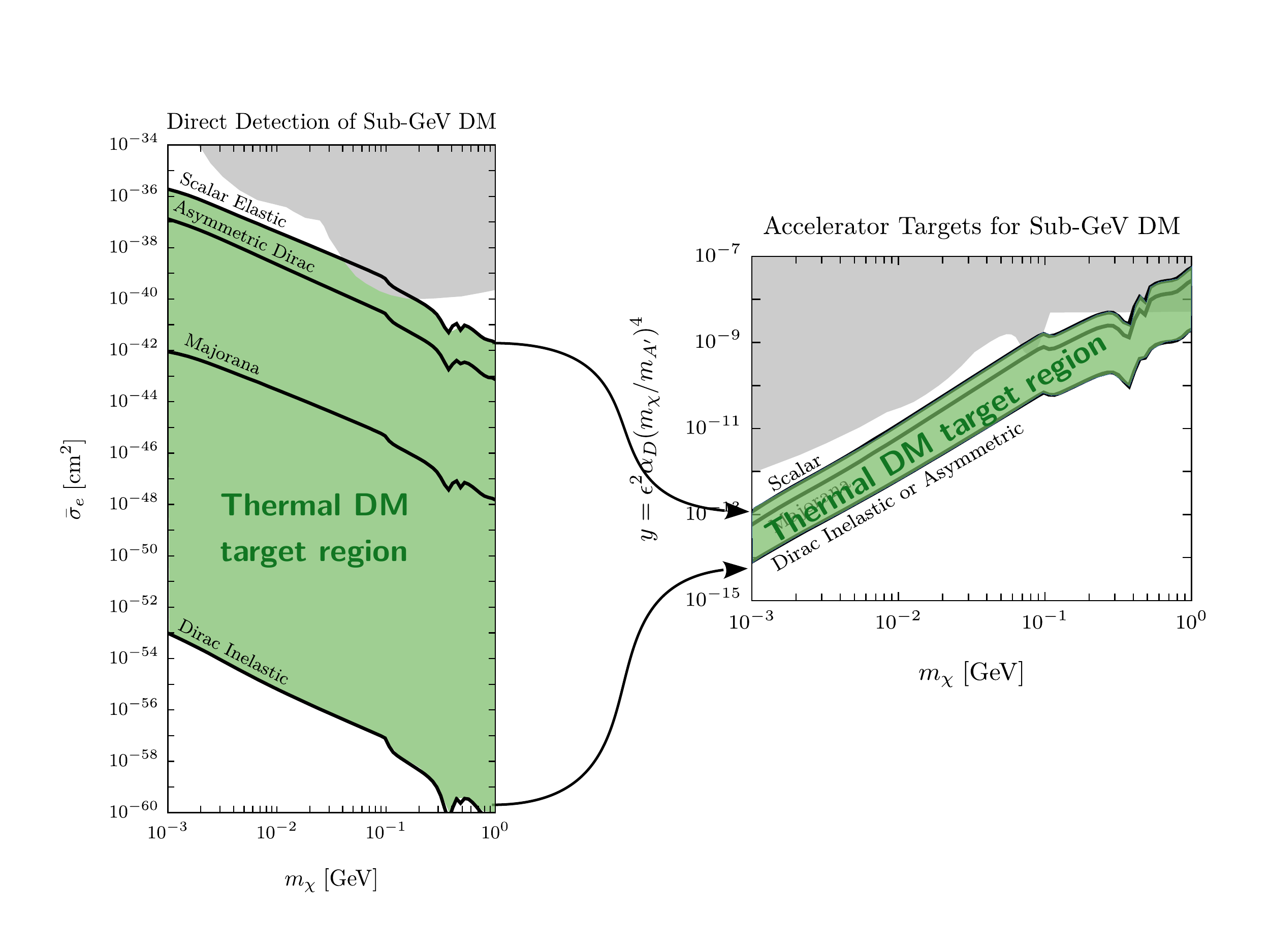}
    \end{center}
    \caption{
    Comparison of sub-GeV DM thermal production targets in the direct detection plane in terms of the electron cross section (left) and on the accelerator plane in terms
    of the variable $y$ (right). Since 
    accelerator production mimics the relativistic kinematics of the early universe, the corresponding signal strength is never suppressed by velocity, spin,
    or small degrees of inelasticity, so the 
    targets are closer to experimentally accessible regions of parameter space. Note, however, that direct detection sensitivity has a complementary
    enhancement for DM candidates with Coulombic
    interactions, which are enhanced at low velocity. 
    \label{fig:collapsing}}
\end{figure}

Accelerator-based light DM searches are highly complementary to another promising avenue for discovery of light DM: low-threshold direct detection (discussed in \cite{CF1_Essig:2022dfa}).  Both approaches are essential to a strong light DM search program.  
There are key differences in what properties they probe --- accelerator-based experiments directly characterize the particle properties of produced DM, while direct detection explores a combination of these properties with their cosmological abundance. They also probe DM in vastly different kinematic regimes: whereas direct detection probes very non-relativistic scattering, accelerators explore relativistic DM production.  This kinematic difference translates into complementary discovery potential: low-threshold direct detection is a uniquely powerful probe for Coulomb-like interactions with enhancements at low velocities, including models of freeze-in through a light mediator.  Accelerators are optimal for discovery of DM whose interactions are suppressed at low velocities, including thermal freeze-out through a dark photon with generic spin and mass structure. 
As shown in Fig \ref{fig:collapsing}, depending on the Lorentz structure of the
of the dark-visible interaction, the non-relativistic direct detection cross section can be suppressed
by many orders of magnitude while relativistic accelerator production is not suppressed by such variations.
Still other thermal (and non-thermal) models, including elastically interacting scalar DM, can be observed efficiently by both approaches, allowing exciting opportunities to characterize any observed signal. 

\subsection*{Science Opportunities and The Road Ahead}
\begin{figure}[h!]
\begin{center}
\includegraphics[width=0.64\hsize]{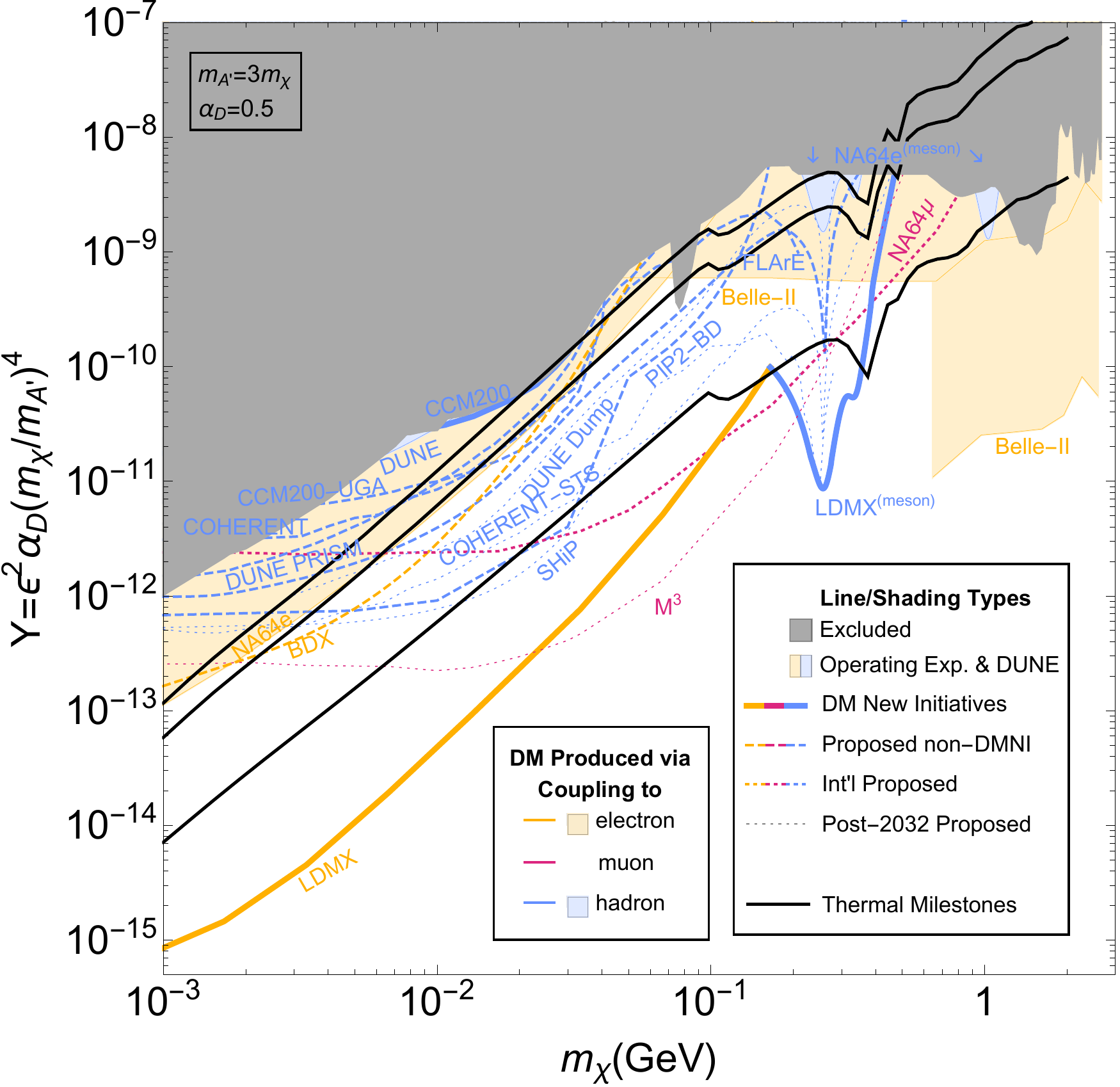}
\end{center}
\caption{
\label{fig:execDarkPhoton}
Thermal milestones for the kinetically mixed dark photon model, a key light DM benchmark (shown as black solid lines), along with exclusions from past experiments (gray shaded regions), projected sensitivities of future projects that are operating or have
secured full funding (colored shaded regions), and projections for proposed experiments (unfilled curves).  The (completed or proposed) experiments selected by Dark Matter New Initiatives (DMNI) program are shown by solid colored lines; other proposed experiments and upgrades that can be realized within a decade are shown as long-dashed colored lines if they are based in the US and/or have strong US leadership, or as short-dashed colored lines if they are primarily international efforts. Proposed experiments that are farther into the future are shown as thin dotted lines.  As can be seen, the combination of operating experiments and the DMNI pre-project experiment (LDMX) can fully explore the thermal DM targets, considered a major milestone of dark-sector physics. 
Each line/region is color-coded according to which SM fermion coupling is employed to produce the DM. In variations of this model, some species may have suppressed couplings, making lines of different colors complementary in this extended model space.
The figure focuses on a parameter slice with mediator mass fixed to 3$\times$ the DM mass, and dark-sector coupling $\alpha_D=0.5$.   As these parameters are varied, the thermal milestones stay approximately fixed in the $y$ vs. $m_\chi$ plane (except in a near-resonant annihilation region) while experimental sensitivities generally improve, as illustrated for a subset of experiments in Fig.~\ref{sfig:darkPhotonThermalVaryR}.
 }
\end{figure}

In the past decade, a key goal of the light DM search effort has been broadly exploring DM models in the MeV to GeV mass range. The simplest, and most WIMP-like, viable mechanism for light DM thermal freeze-out is annihilation to SM particles via an $s$-channel dark photon.  This model has therefore emerged as a key benchmark model.  Because DM production at (semi)-relativistic kinematics drives both the dynamics of freeze-out and DM production at accelerators, the range of freeze-out interaction strengths (often parametrized by a dimensionless parameter $y$ related to the effective Fermi scale of the interaction) compatible with this mechanism is narrow, spanning a factor of $\sim 30$ at a given DM mass (black diagonal lines in Fig. \ref{fig:execDarkPhoton}). 

This milestone was identified as a high-priority goal for the accelerator-based program by the Dark Matter New Initiatives (DMNI) BRN workshop \cite{BRN} and  subsequent summary 
\href{https://science.osti.gov/-/media/hep/pdf/Reports/Dark_Matter_New_Initiatives_rpt.pdf}{report}.  
Following this, a competitive DMNI process by DOE HEP selected two intensity-frontier projects to support, CCM200 and LDMX, to explore this milestone with different timescales and complementary sensitivity.  CCM200, a proton beam re-scattering experiment at Los Alamos' LANSCE, was completed and commissioned in 2021 and is now operating.  LDMX, a missing momentum experiment at SLAC's LESA electron beamline, received pre-project funds, awaits construction funding, and could begin operation in FY26. CCM200 expands sensitivity to hadronic DM couplings, while LDMX will explore all thermal DM milestones below $\sim 1/2$ GeV, complementing the sensitivity of Belle-II to $\sim$ GeV mass thermal DM milestones. These experiments' sensitivity projections are illustrated in Fig.~\ref{fig:execDarkPhoton}. 
These experiments' coverage of thermal milestones is robust to many important model uncertainties, such as varying dark-sector couplings and the DM to dark-photon mass ratio (excepting a fine-tuned resonance-enhanced region). 

Fig.~\ref{fig:execDarkPhoton} and this report also highlight the DM-search capabilities of  experimental concepts beyond those selected for DMNI funding.  As detailed in \cite{Ilten:2022lfq}, most of these experiments can also be realized within the next decade, if supported for example by subsequent rounds of DMNI pre-project and construction funding.  The breadth of ideas within this program is valuable for several reasons.  The use of multiple complementary techniques will assure a robust program, and in the case of discovery the ability to measure dark sector masses and interaction strengths. 
Multiple, complementary experiments are also important to probe generalizations of thermal freeze-out. Some of these, such as those where a mediator does not couple to electrons but preferentially to $\mu$ and/or $\tau$ leptons or baryons, motivate a continuing push in missing-X and re-scattering experiments to improve sensitivity to muon and hadron coupled DM production (echoing ``Thrust 1'' of the accelerator Priority Research Direction in \cite{BRN}).  Others, including models where meta-stable particles in the dark sector play important roles in DM cosmology and enable new discovery techniques, and neutrino portal annihilation with qualitatively different experimental signals, motivate searches for semi-visible DM signals and DM-motivated visible signals (echoing ``Thrust 2'' from \cite{BRN}). We note that DMNI has not yet funded any experiments optimized for (semi)visible dark sector searches, but the next generation of small-project proposals could cover substantial parameter space with strong DM motivation, again complementing Belle II and LHCb capabilities.

These motivations and corresponding experimental opportunities are 
examined in more detail in Figs.~\ref{fig:globalVectorsThermal} to \ref{fig:semiVisible}.  Related techniques can also advance the detection of millicharged particles (Fig.~ \ref{fig:millicharge}), which present a distinctive detector signature and could make up a small fraction of the DM.  Finally, theory has a key role to play in defining the future of the light DM search program --- both by continuing to explore the space of light DM models and through theory-experiment collaborations, which have played an important role in the development of many of the concepts and analyses considered here.


The field of concepts in intensity-frontier experiments searching for DM has grown tremendously in the last decade, in response to the tremendous untapped discovery opportunities that it presents. Most of these concepts are low-cost, based on either analyses of multi-purpose experimental data or small experiments that leverage existing accelerator infrastructure and detector technologies.  In the next decade, the realization of this opportunity through funding for dark-sector searches, completion of the DMNI-supported program, and the selection of complementary concepts in subsequent round(s) of DMNI will shed a clear light on the possibility of low-mass particle DM and other light new physics.

\section{Introduction}

The existence of dark matter (DM) is clear evidence of physics beyond the 
Standard Model (SM), yet its particle identity remains elusive despite
decades of experimental searches primarily aimed at TeV scale weakly interacting
massive particles (WIMPs) \cite{Bertone:2016nfn}.
In recent years, it has become widely appreciated that SM neutral DM
in the MeV-GeV range can realize many compelling and predictive
cosmological production mechanisms by interacting  with the SM through light 
new force carriers (``mediators"). In parallel, there has also been a fertile effort to
design new searches for such particles at existing and proposed high-intensity
accelerator facilities \cite{Battaglieri:2017aum}. In this document, we survey
the landscape of such efforts with an emphasis on dark matter production and its
subsequent detection in a controlled laboratory environment.

\begin{figure}[hbt!]
    \includegraphics{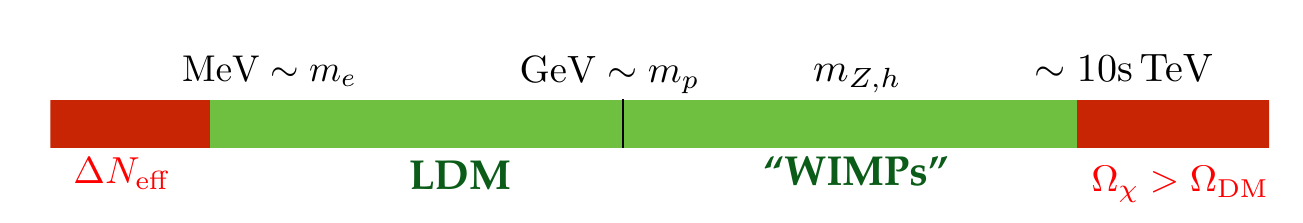}
    \caption{Viable mass range for DM that
    equillibrates with visible
    matter in the early universe.
    Masses below $\sim$ MeV conflict 
    with BBN  \cite{Nollett:2013pwa} and masses above 
    few 10s of TeV violate perturbative
    unitarity for interaction strengths that avoid cosmological overproduction \cite{PhysRevLett.64.615}. This mass range applies
    to any scenario in which DM is thermalized with SM particles prior to BBN, independently of how its late-time abundance is ultimately set. }
    \label{fig:my_label}
\end{figure}

 Unless the DM-SM interaction is 
{\it ultra feeble} \cite{Hall:2009bx}, the particles in the dark sector will generically be in chemical equilibrium 
with visible matter in the early universe. As a corollary, every DM candidate testable at 
accelerators is necessarily in equilibrium at early times, under standard
cosmological assumptions. Independently
of how its excess thermal entropy is depleted to achieve the 
observed DM density, any equillibrated DM candidate 
must have mass in the $\sim$ MeV-100 TeV range; sub-MeV
thermalized particles conflict with 
cosmological observables \cite{Nollett:2013pwa} and particles above the 
$\sim 100$ TeV scale with sufficient interaction strength violate 
unitariy \cite{PhysRevLett.64.615}.
The upper (10 GeV-100 TeV) half of this viable range is currently being probed
by nuclear-recoil direct detection, indirect detection, and 
high energy collider searches. By contrast, the lower (MeV-10 GeV)
range of the thermal window is comparatively under-explored with
few dedicated experiments for light DM.

Unlike heaver WIMP-like DM, which can undergo freeze out through the SM weak force, 
sub-GeV DM must couple to light new mediators to yield the observed abundance; either
through canonical thermal freeze out or via alternative mechanisms~ \cite{Lee:1977ua,Boehm:2003hm}; for such light states interactions
through weak scale (or heavier) particles violate the Lee-Weinberg bound and overclose the universe.
If freeze out is a one-step process that directly annihilates the DM into SM particles (as opposed
to annihilation to new unstable particles), the DM-SM coupling is in one-to-one correspondence
with the DM abundance and sets a sharp experimental target for accelerator searches. 

Existing experimental limits require sub-GeV DM particles to be neutral under SM gauge interactions;
 any such states would have
been discovered at previous colliders \cite{Egana-Ugrinovic:2018roi}. Consequently,
mediator interactions with the SM must proceed
through renormalizable ``portals" (SM singlet operators of mass dimension $<$ 4) and there is a 
finite list of such portals 
\begin{equation}
\label{portals}
F_{\mu\nu}~~,~~ LH~~,~~ H^\dagger H~~,~~ \bar f \gamma^\mu f~~,
\end{equation}
where $F_{\mu\nu}$ is the tensor ``vector portal" that enables kinetic mixing between the photon and other vectors, $LH$ is the ``neutrino portal" that enables mixing between 
neutrinos and other neutral fermions, 
$H^\dagger H$ is the Higgs portal that 
enables Higgs mixing with new scalar mediators,  and $\bar f \gamma^\mu f$ is a SM current (e.g. $B-L)$ that can couple to the vector
boson of a new $U(1)$ gauge extension to the SM. Thus, in contrast with
heavier scenarios, light DM models offer a finite list of possible interactions with 
the SM, which sharpens the focus of the experimental discovery effort.
While light mediators can also couple to non-renormalizable SM singlet operators,
such interactions would be insufficient to 
evade the Lee-Weinberg bound \cite{Lee:1977ua}. 

For each portal in Eq.~\ref{portals} the corresponding mediator can also couple to DM
and mediate thermal relic annihilation such that the DM abundance is in one-to-one correspondence with
the mediator's coupling to visible matter. Furthermore, since the mediator's coupling to DM must satisfy upper limits from perturbative unitarity \cite{PhysRevLett.64.615}, ensuring a sufficiently large cross section for thermal freeze out implies  in  there is a {\it minimum} mediator coupling to visible matter in each scenario. Thus, reaching sensitivity to
these minimum couplings can discover or falsify a broad range of predictive DM models below the GeV scale.

Current and future high-intensity accelerators offer  essential discovery modes for light DM with sensitivity to key
milestones for early universe production. These facilities include
proton, electron, and muon beam fixed target 
experiments, low-energy colliders, and forward positioned LHC detectors. Such facilities
enable three basic approaches to detection:
\begin{itemize}
    \item {\bf Rescattering:} Accelerator produced DM particles can scatter in a downstream detector.
    \item{\bf Missing Energy/Momentum Signals:}
    Dark matter particles produced in in an accelerator yield missing energy and momentum in signatures
    \item{\bf Semi-Visible Signatures:} In many dark sector models, transitions between dark sector
    states (e.g., via decays) can produce additional visible particles 
\end{itemize}
{\bf A mature program that combines these accelerator based efforts suffices
to cover nearly all of the predictive
thermal targets based on direct annihilation. 
}

Accelerator searches are also complementary to 
the new program of low-mass direct detection
techniques based on non-traditional 
targets (e.g. electrons). 
Such efforts also can also probe predictive thermal targets in models that predict velocity unsuppressed elastic scattering in the non-relativistic limit, but are insensitive to the thermal targets
in models with velocity suppression or inelastic
kinematics.
By contrast, accelerator production recreates
the kinematic conditions of the early universe and 
is unaffected by the spin of the DM or the lorentz structure of its interactions in this regime. 
{\bf Thus, a broad experimental program encompassing both accelerator and 
direct detection searches is necessary to fully 
test the thermal origin of light DM and determine
its detailed properties in the event of a discovery.
}


\section{Approaches to Dark Matter Searches}
This section reviews three main classes of accelerator-based searches for dark matter. These are primarily distinguished by the signatures they use to detect DM production: 

\begin{enumerate}
    \item  \textbf{Missing Energy, Momentum, or Mass Searches} exploit the kinematics of recoiling particles from the dark matter production vertex, and appropriate vetoes, to identify DM production events. These strategies require a sensitive detector with a direct view of the interaction region, and typically benefit from relatively low luminosities to allow clean vetoes of Standard Model background reactions.
    
    \item \textbf{Re-Scattering Searches} aim to detect the scattering of DM particles or millicharges in a detector volume. These require the highest luminosities of all experiments we consider (because the detection efficiency per DM particle produced is parametrically small) and rely on material between the interaction region and detector to shield the latter from SM interaction products.  Though conceptually different, each of these approaches offers inclusive sensitivity to \emph{any} dark matter candidate that interacts sufficiently to be produced in SM interactions (as is guaranteed --- at least for some species --- if DM is produced through the most direct thermal freeze-out mechanisms).
    
    \item \textbf{Searches for Semi-Visible Signals of DM Production} explore the possibility that DM is the stable matter of a larger dark sector, whose unstable particles decay into a combination of DM and SM particles. Spectrometer-style and displaced-decay experiments can also offer a powerful window on these scenarios for dark matter production\footnote{Semi-Visible Signals have cross-cutting connections with several other RF6 whitepapers. As signals of both Dark Matter Production \emph{and} Extended Dark Sectors, they are also discussed both here and in \cite{BI3}. In addition, experiments designed to probe ``Minimal Portals to Dark Sectors'' \cite{BI2} are often well equipped to perform these searches.}.
\end{enumerate}  

Detailed discussion of specific experiments and facilities will be deferred to \cite{RF6ExpFacilities}. Our aim here is instead to discuss the range of conceptual approaches, highlighting both their common features and their complementarity.  We mention specific experiments simply to assist the reader in relating our discussion to \cite{RF6ExpFacilities}, and also refer the reader to that whitepaper for references to experimental proposals. 


\subsection{Missing Mass/Energy/Momentum Searches at Colliders and Fixed-Target Facilities} 
If all visible final-state particles' momenta are reconstructed, and the initial-state 4-momentum is known, then energy and momentum conservation can be used to infer the invariant mass of all invisible particles.  This \emph{missing mass} may be peaked at the  mediator mass if it can be produced on-shell, or for off-shell mediators has a threshold at twice the DM mass.   The strict prerequisites for this kind of measurement are satisfied for a limited class of initial states and experimental configurations, such as $e^+e^-$ annihilation (at a collider such as Belle II or with positron beam impinging on a thin target where it can annihilate with atomic electrons) or meson decays (when the 4-momentum of a daughter meson can be inferred from a tagged parent meson and its visible decay products, as discussed e.g.~in \cite{LOI_invis_eta}).  

In many lepton-beam fixed-target experiments, unobserved recoiling nuclei in the final state prevent the full reconstruction of the missing 4-momentum. However, these nuclei generally carry modest momentum and (because they are very non-relativistic) very little energy, so that the energy of all non-nucleus products is, to high accuracy, equal to the beam energy.  Furthermore, when DM is produced via dark bremsstrahlung or $e^+e^-$ annihilation, it typically carries the majority of this energy. Thus, dark matter events can be kinematically characterized by the calorimetric ``disappearance'' of a sizable fraction of the beam energy, and a further veto on any hadrons or muons that could carry this energy but not deposit it in a calorimeter.  This \emph{missing energy} signal \cite{Andreas:2013lya} is the foundation of the NA64 concept.  A \emph{missing momentum} experiment \cite{Izaguirre:2014bca} is a refinement of this approach, separating the interaction region from the calorimeter so that the recoil lepton from dark bremsstrahlung can be identified and its transverse momentum measured.  This enables rejection of charged-current backgrounds, an additional handle for rejection of photon-initiated backgrounds, and discrimination of the mass scale of produced DM pairs~\cite{Blinov:2020epi} (in addition to allowing a missing-energy analysis of events where the beam particle enters the calorimeter with full energy). Both of these approaches can be realized using electron, positron, or muon beams.
 In \emph{positron missing energy} experiments such as the POKER initiative, where a positron beam impinges on a thick target, a resonance in $e^+e^-$ annihilation to DM \cite{Marsicano:2018glj} will lead to a missing energy peak at $E_{miss} = m_{res}^2/{2 m_e}$.  This technique therefore offers excellent prospects for measuring the mass of dark-sector particles that decay into DM, provided they are kinematically accessible. 
 Similar approaches may also be realized based on dark Compton scattering in photon beams \cite{Chakrabarty:2019kdd,LOI_compton}.
 
 Although hadron beams are not conducive to ``missing-X'' experiments, mediators with dominantly hadronic couplings can still produce a missing energy signal in $e^\pm$ beam experiments via exclusive meson photo-production, with subsequent rare meson decays to DM through mixing with the hadronic mediator \cite{Schuster:2021mlr}, as well as missing-mass signals in meson decays.

\subsection{Re-Scattering}
Instead of detecting the DM production event, \emph{DM re-scattering} experiments rely on a high luminosity of fixed-target or collider reactions to produce a secondary beam of DM particles, then detect their scattering in a forward detector \cite{Batell:2009di}.  This scattering proceeds through exchange of a mediator with very weak SM couplings, and so the typical DM particle does not interact even in a large detector volume. Therefore, sensitive experiments must rely on intense beams or high-luminosity colliders.

This concept has been realized to date by using a primary proton beam to produce DM, and detecting it in neutrino detectors such as LSND \cite{Batell:2009di}, MiniBooNE \cite{MiniBooNEDM:2018cxm}, COHERENT \cite{COHERENT:2019kwz}, or CCM \cite{CCM:2021leg}.  Future proposals will improve on these experiments' sensitivities with higher beam energies (improving sensitivity to higher-mass DM), and increased beam intensities, detector thresholds, and/or detector masses (improving sensitivity to weaker DM couplings) as well as reducing detector and neutrino backgrounds.  Proton-beam experiments are highly synergistic with the accelerator-based neutrino physics program because they use the same beamlines and detectors, sometimes with dedicated running configurations. 
DM re-scattering at electron beams \cite{Izaguirre:2013uxa}, as proposed by BDX, has the advantage of a more compact secondary DM beam, but requires a dedicated detector downstream of a high-luminosity electron fixed-target facility such as CEBAF (or in the future, a high-energy linear collider). 

Recent calculations have shown that $pp$ collisions at the LHC can also produce a high luminosity of low-$p_T$ DM particles in the forward direction, detectable by a similar approach in the Forward Physics Facility (FPF)~\cite{Anchordoqui:2021ghd,Feng:2022inv} in experiments such as FLArE~\cite{Batell:2021blf}. LHC collisions are particularly advantageous for exploring GeV-scale DM masses through their nuclear scatterings, while can also allow for electron scattering searches at lower DM masses~\cite{Batell:2021aja}. In this case, the specific far-forward location and the impact of the LHC infrastructure allows for efficient reduction of low-energy neutrino-induced backgrounds in such searches.

Finally, several groups are pursuing the use of both proton beam dumps and LHC collisions for \emph{rescattering detection of millicharged particles}.  This signal differs from the DM scattering discussed above in that the scattering occurs through exchange of the photon (or, in the case of millicharge-like dark sector models, an ultra-light dark photon) rather than a massive mediator, so low-momentum-transfer scattering are enhanced. As such, millicharges undergo repeated, very soft scatters in material, leading to a very low $dE/dx$ rather than individual hard scatters with $\gtrsim$ MeV momentum transfer. Proposals in this class include (alphabetically) FerMINI~\cite{Kelly:2018brz}, FORMOSA~ \cite{Foroughi-Abari:2020qar}, MilliQAN~\cite{Haas:2014dda}, and SUBMET~\cite{Choi:2020mbk}. 

Though this whitepaper is focused on light DM from a hidden sector, we note that conceptually similar approaches are relevant to other models of light dark matter.  For example, a DM particle with long hadronic interaction length, such as stable sexaquarks, can also be probed for with a re-scattering  experiment concept that shares some commonalities with both millicharge and DM-rescattering scenarios discussed above \cite{Farrar:2022mih,LOI_sexaquark}.  

\subsection{Semi-Visible Signals}
\label{ssec:semivisible}
Previous examples focused on the production and detection of the stable DM particles that arise in minimal dark sector models. In certain scenarios, however, the dark sector also contains \emph{unstable} particles that participate in the early-universe thermal mechanisms responsible for the production of DM.  We will refer to these collectively as DM excited states.  In these models, the excited states can be produced in an accelerator-based experiments and decay to the stable DM and SM particles after traversing a macroscopic distance in the detector. This leads to a displaced ``semi-visible'' signal where part of the excited state energy and momentum is carried away by the DM and part is deposited in the detected SM particles. Remarkably, even these rich dark sectors can feature concrete theoretical targets --- as will be illustrated in Sec.~\ref{ssec:iDMandSIMPs} below. Searches for semi-visible signals can often be performed as variations on searches for minimal dark sector portals (as discussed in another RF6 whitepaper \cite{BI2}) by experiments such as DarkQuest \cite{Apyan:2022tsd}, Heavy Photon Search \cite{Baltzell:2022rpd}, or FASER \cite{Berlin:2018jbm}, as a perturbation of ``Missing X'' searches (for example at Belle II \cite{Duerr:2020muu}), or in detectors designed for re-scattering \cite{Izaguirre:2014dua}.  

\section{Models and Frameworks for Light Dark Matter and Their Experimental Prospects}

\subsection{Thermal Production and Parameter Space Milestones: Generalities}

In robustly predictive models of thermal freeze out, the DM abundance is 
set by direct annihilation to SM particles via $s$-channel mediator exchange as the early universe cools
 to temperatures just below the DM mass. 
 For any $2\to 2$ reaction in the non-relativistic regime, to
  within order-one factors, the 
 annihilation cross section generically scales as 
 \begin{equation}
    \langle \sigma v \rangle_{\rm ann} \propto 
     g_{\chi}^2 g_{\rm SM}^2\left( \frac{m_\chi^2}{~ m_{\rm med}^4}\right)~~,
 \end{equation}
 where $g_\chi$ is the mediator-DM coupling, $g_{\rm SM}$ is the mediator-SM coupling, $m_\chi/m_{\rm med}$ is the DM-mediator mass
ratio. It is convenient to define a 
 dimensionless variable
 \begin{equation}
\label{eq:y}
y \equiv g_{\chi}^2 g_{\rm SM}^2\left(\frac{m_\chi}{m_{\rm med}} \right)^4 ~~,~~\langle \sigma v \rangle_{\rm ann} \propto \frac{y}{m_\chi^2}~~.
 \end{equation}
 Away from special kinematic regions (e.g. the $m_{\rm med} \approx 2 m_\chi$ resonance) the variable $y$ is insensitive to the relative values of 
the input parameters, and for each value of $m_\chi$ there is a 
unique prediction for $y$ that sets the observed relic density.
{\bf Reaching experimental sensitivity to these milestones
can unambiguously discover or falsify predictive dark matter
candidates with a thermal origin}

The parametrization in Eq.~(\ref{eq:y}) is broadly applicable
across models of light thermal DM and defines a sharp thermal production milestone in the $y$ vs. $m_\chi$ plane for a given choice of DM and mediator particles. The mediators 
can be any SM singlet particle that couples
linearly to one of the portals in Eq.~(\ref{portals}).
Unlike heavier WIMPs, viable models of light DM 
require suppressed annihilation to electromagnetically charged matter during the CMB era \cite{Planck:2018vyg}, so there are no observable indirect detection signals in these models; new techniques play a crucial role in the 
sub-GeV discovery effort.

In contrast with direct detection, dark matter production at accelerators
is highly relativistic and recreates the kinematic conditions 
of the hot early universe. Thus, variations in the lorentz structure
of the DM-SM interaction yield only order-one differences
in accelerator signal rates, so each search strategy is generically
sensitive to broad variety of models. This feature complements
direct-detection searches whose sensitivity depends crucially
on the spin and momentum dependence of the DM-SM interaction in the
non-relativistic limit. In the event of a discovery, the interplay
of accelerator and direct detection results would provide essential
clues about the identity of the dark matter and the nature of its
its non-gravitational interactions with visible
matter. 

\subsection{Dark matter production through the vector portal} 

Among the most commonly studied mediators is the ``dark photon" $A^\prime$, a gauge boson of a dark $U(1)_D$ group under which SM
 particles carry no charge. The $A^\prime$ can kinetically mix through the vector portal, $\epsilon F^{\mu\nu} F^\prime_{\mu\nu}$ which induces
 an $A^\prime$ coupling to SM fermions
 \begin{equation}
 {\cal L}_{\rm int} = \epsilon e  A^\prime_\mu
 J_{\rm em}^\mu,
 \end{equation}
 where $J^\mu_{\rm em}$ is the 
 electromagnetic current and $\epsilon \ll 1$ is the kinetic mixing parameter whose naturally
 small magnitude can arise from integrating out heavy new particles charged under 
 $U(1)_D$ and $U(1)_Y$. 
 Depending on the details of $U(1)_D$ breaking,
 the dark photon mass $m_{A^\prime}$ can
 have any non-negative value. However, 
 dark matter with long range forces and appreciable couplings are subject to stringent
 limits from DM self interactions \cite{Tulin:2017ara}. In the massless $A^\prime$
 limit, particles charged under $U(1)_D$
 acquire SM millicharges, which can only
 constitute all of the DM for extremely
 feeble couplings not suitable for accelerator production; such particles with 
 larger couplings can be a subdominant
 fraction of cosmological dark matter and 
 also be produced in accelerators (see Sec. \ref{sec:millicharges}). 
 
 For a model to realize a predictive and testable thermal history, the relic density must depend on the 
 mediator coupling to visible matter. This is guaranteed if the dominant annihilation
 process is $s$-channel 
 $\chi\chi \to f^+ f^-$ annihilation, where $f$ is
 a charged SM particle and we demand $m_\chi < m_{A^\prime}$ to ensure that the less predictive \footnote{However, see Ref. \cite{DAgnolo:2015ujb} for an exception near
 $m_\chi \lesssim m_{A^\prime}$.} $\chi \chi \to A^\prime A^\prime$ process
 does not contribute. Furthermore,
 satisfying CMB limits on energy injection after recombination 
 \cite{Planck:2018vyg}
 requires that the annihilation rate is suppressed
 after freeze-out (i.e. either the cross section
 is $p$-wave or the DM population changes
 after freeze-out). 
 
 Here we present three representative DM candidates that viably realize $s$-channel annihilation
 through the vector portal:
 
 \begin{itemize}
     \item {\bf Complex Scalar:} the mediator DM interaction is 
     \begin{equation}
         {\cal L}_{\rm int} = g_D A^\prime_\mu \chi^* \partial^\mu \chi + h.c. \label{eq:scalar}
     \end{equation}
     where $g_D \equiv \sqrt{4\pi \alpha_D}$ is the dark coupling constant.
     Here $\chi \chi^* \to f^+f^-$ annihilation is $p$-wave 
     and safe from CMB limits and this model is testable with
     both accelerator probes and direct detection
     experiments sensitive to sub-GeV DM (e.g. via electron recoils)
     \item {\bf Majorana Fermion:} the mediator DM interaction is 
     \begin{equation}
         {\cal L}_{\rm int} = \frac{g_D}{2} A^\prime_\mu \bar \chi \gamma^\mu\gamma^5 \chi, \label{eq:majorana}
     \end{equation}
     where the $s$-channel annihilation is also $p$-wave 
     and safe from CMB limit. Here the non-relativistic 
     direct-detection cross section is suppressed by 
     $v^2 \sim 10^{-6}$ in the Galactic halo, so direct detection
     sensitivity is sharply diminished for this lorentz structure,
     but relativistic accelerator production is unaffected. 
     
      \item {\bf Pseudo-Dirac:} Here the dark sector consists
      of two Weyl fermions with an appreciable Dirac mass and small Majorana masses; the hierarachy of masses is natural as the Majorana masses break the dark sector's analogue of lepton number. In the mass basis, the two
       eigenstates $\chi_{1,2}$ have a small mass
       splitting $\Delta \equiv m_2- m_1$, and couple off-diagonally to the mediator
     \begin{equation}
         {\cal L}_{\rm int} = g_D A^\prime_\mu \bar \chi_1 \gamma^\mu \chi_2 + h.c. \label{eq:pseudoDirac}
     \end{equation}
     Here the relic density is set by $\chi_1 \chi_2 \to f^+f^-$ {\it coannihilation} and the heavier state $\chi_2$ is unstable and becomes depleted via
     $\chi_2 f \to \chi_1 f$ scattering of $\chi_2 \to \chi_1 f^+ f^-$ long after freeze-out. Thus, the model is 
     safe from CMB bounds and inaccessible via direct detection due to the inelasticicty of the interaction \cite{CarrilloGonzalez:2021lxm}
     
 \end{itemize}
 
 In each of these models, the dimensionless interaction
 strength $y$ in Eq.~(\ref{eq:y}) can be written more concretely as 
 \begin{equation}
     y \equiv \epsilon^2 \alpha_D \left( \frac{m_\chi}{m_{A^\prime}}\right)^4,
 \end{equation}
where the specific value of $y$ required for freeze-out
depends only mildly on the model variations.

\begin{figure}[htbp!]
\begin{center}
\begin{subfigure}[t]{0.6\hsize}
\includegraphics[width=\hsize]{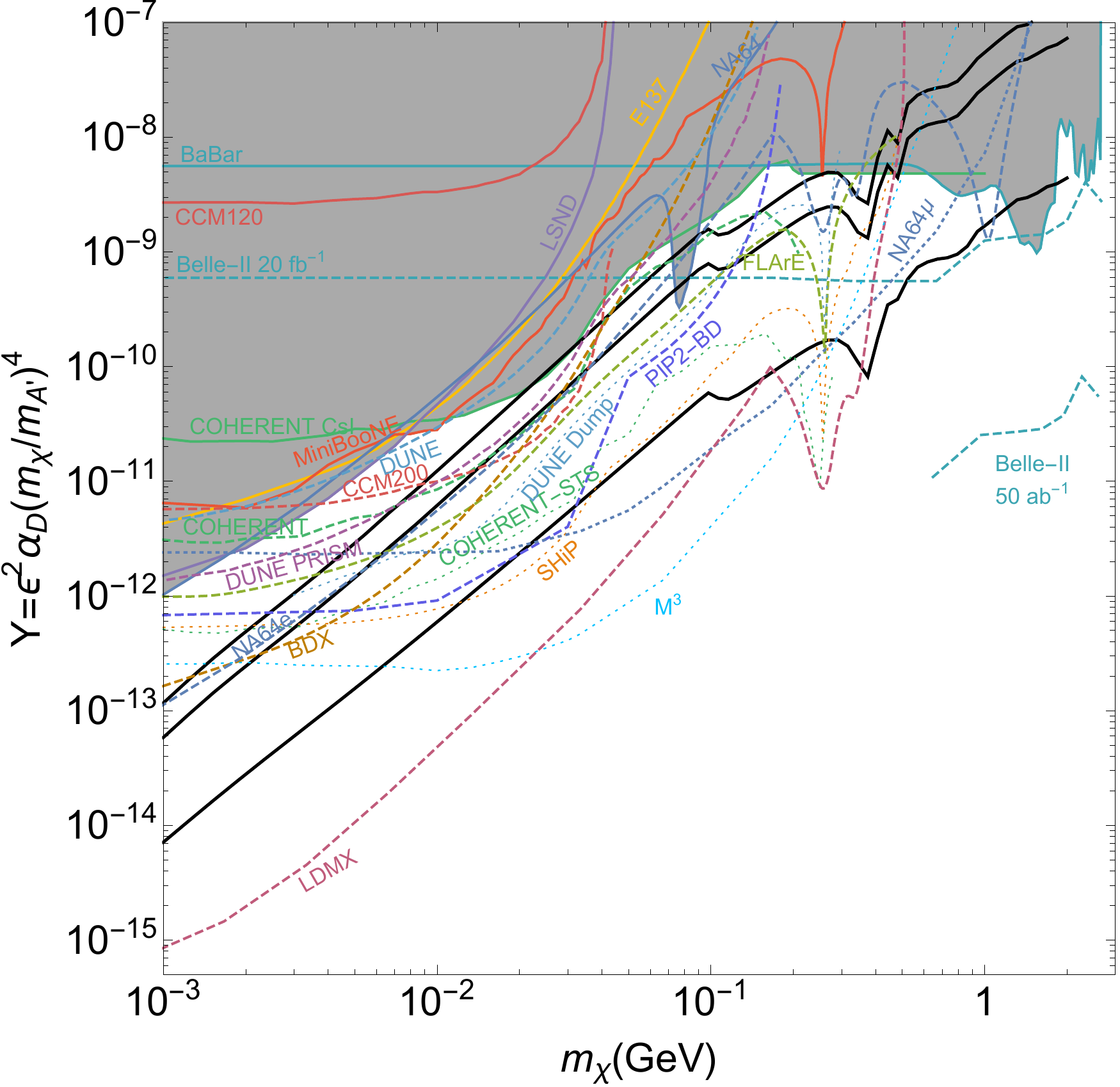}
\caption{Thermal Milestones and Experimental Sensitivities}
\label{sfig:darkPhotonThermalR3}
\end{subfigure}
\\
\begin{subfigure}[t]{0.7\hsize}
\includegraphics[width=\hsize]{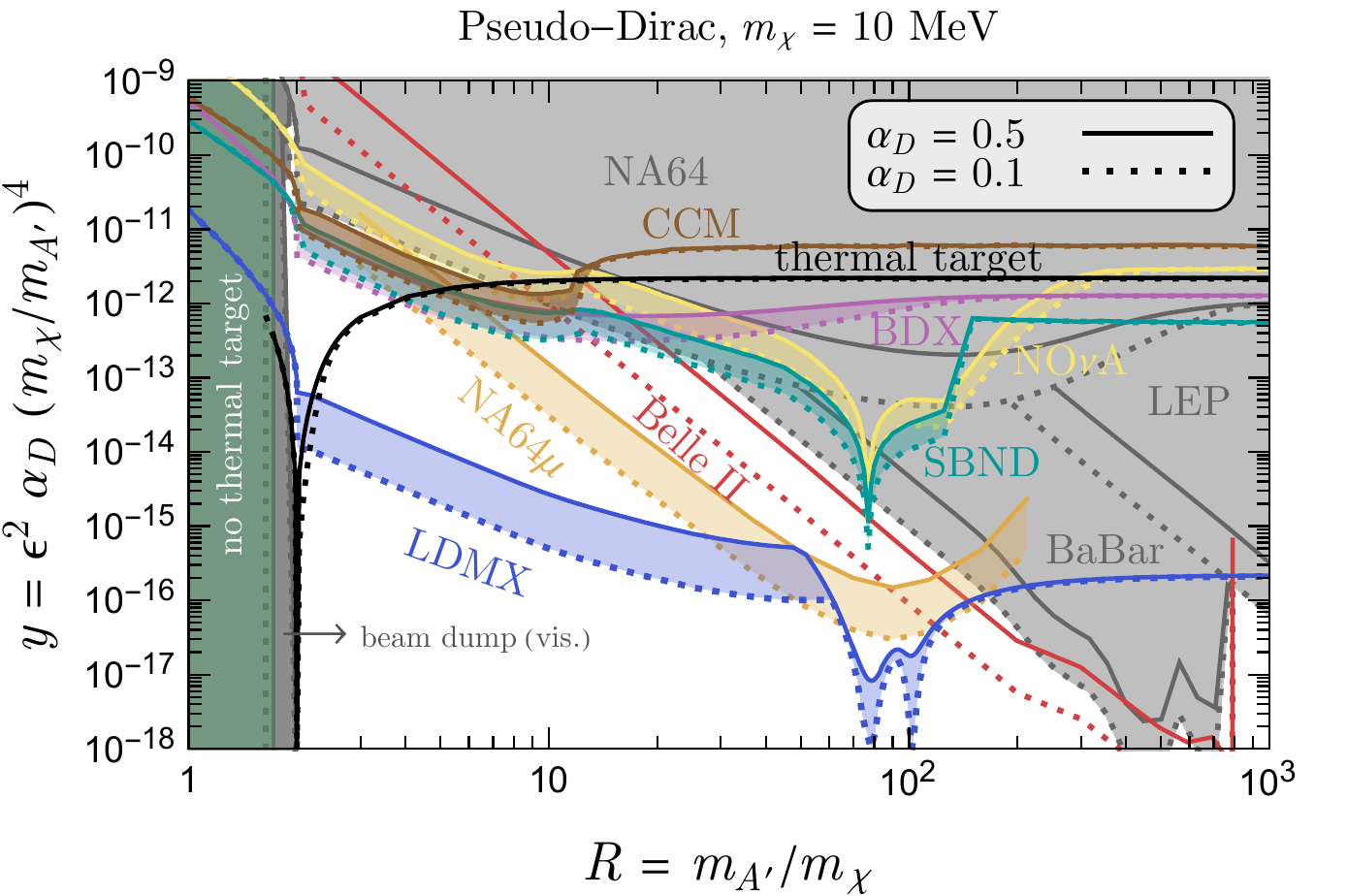}
\caption{Variation of Sensitivity and Thermal Milestone with $R=m_{A'}/m_\chi$ and $\alpha_D$, with $m_\chi=10 \MeV$ fixed.}
\label{sfig:darkPhotonThermalVaryR}
\end{subfigure}
\end{center}
\caption{\subref{sfig:darkPhotonThermalR3}:
Kinetically mixed dark photon targets for various DM spins with DM/mediator mass ratio 3 and coupling $\alpha_D = 0.5$. \subref{sfig:darkPhotonThermalVaryR} illustrates variation of exclusions and projected sensitivities (for an illustrative subset of experiments of different types) for pseudo-Dirac DM as a function of $R = m_{A'}/m_\chi$, with $m_\chi$ fixed to $10 \MeV$, and $\alpha_D=0.5$ (solid) and 0.1 (dashed).  Mass-dependence and methodology are further discussed in \cite{Berlin:2020uwy}.
\label{fig:darkPhotonThermal}}
\end{figure}

\subsubsection{Parameter Variations}


The thermally motivated target for $y$ is approximately only a function of the DM mass $m_\chi$, provided that the freezeout of $\chi \bar{\chi} \to A^{\prime *} \to f^+ f^-$ in the early universe occurs away from kinematic regions in which $A^\prime$ can be produced on shell, i.e., provided $m_{A^\prime} \not\approx 2 m_\chi$ (mass-dependence in this near-resonance region was derived in \cite{Feng:2017drg}). Note that direct searches for new mediators exclude the possibility for very large values of $R \equiv m_{A^\prime} / m_\chi$, since $\eps^2 \propto y \, R^4$ must be unfeasibly large in this case (this statement is closely related to the so-called Lee-Weinberg bound on the mass of light thermal DM). Thus, previous studies have focused most heavily on values of  $R$ that are sufficiently large to admit such predictive targets $y = y(m_\chi)$, yet sufficiently small to not yet be ruled out by terrestrial probes. A common choice that approximately satisfies this criteria is $R = 3$. 

Although the standard choice $R=3$  is reasonable, it is valuable to consider the larger parameter space spanned by this mass ratio. Indeed, although it is commonly thought that the sensitivity of low-energy accelerator searches for DM is predicated on the ability to produce on-shell mediators, which in turn decay to DM particles, this is not the case as we highlight below. In fact, the sensitivity of low-energy accelerators to the direct production of DM particles through off-shell processes often scales in a manner that is favorable when compared to  the scaling of the thermal target.

We illustrate this explicitly in Fig.~\ref{sfig:darkPhotonThermalVaryR}, which shows the thermal target (black), existing bounds (shaded grey), and projected sensitivity (colored lines) of various low-energy accelerator searches in the parameter space spanned by $y$ and $R$, for a fixed DM mass of $m_\chi = 10 \ \text{MeV}$ and two different values of the DM coupling $\alpha_D = 0.5$ (solid lines) and $\alpha_D = 0.1$ (dotted lines). From this figure we can glean three main regions relevant for thermal freezeout and direct searches. For $R \gg 1$, the thermal target is independent of the particular value of $R$. Also, in this regime, the center of mass energy involved in various low-energy accelerator probes is well below threshold for on-shell production of the $A^\prime$. In this case, DM production is well described by a contact operator and proceeds through off-shell $A^\prime $ production. As a result, both the value of $y$ favored by freezeout and the experimental sensitivity to $y$ are independent of the particular value of $R$. However, for much higher energy probes, such as BaBar and LEP, DM production production continues to proceed through on-shell $A^\prime$ production, and hence the rate is solely dependent on $\epsilon^2$. As a result, when recast in terms of $y$, higher energy experiments have increases sensitivity for increasing $R$, provided that the dark photon can be produced on shell. 

For smaller values of $R$, near $R \simeq 2$, freezeout is resonantly enhanced in the early universe, favoring much smaller values of $y$ and introducing a weak dependence on $\alpha_D$ due to the effect of the $A^\prime$ width on the thermally-averaged annihilation rate. Thus, in a tuned range near $R \simeq 2$, terrestrial probes lack sufficient sensitivity to cover the entirety of the thermally motivated region. For $R < 2$, freezeout can efficiently proceed through $\chi \bar{\chi} \to A^\prime A^\prime$, in which case freezeout is independent of the coupling $\epsilon$ to the SM. This is shown as the solid green region of Fig.~\ref{sfig:darkPhotonThermalVaryR}, in which there is no cosmologically favored value of $y$ (aside from the fact that thermalization requires sufficiently large $\epsilon$ and so bounds this coupling from below). Furthermore, for $R \lesssim 2$, DM production must still proceed through off-shell processes since $A^\prime \to \chi \bar{\chi}$ is kinematically forbidden. In this case, DM production is independent of the DM mass, such that the senstivity to $y$ scales as $y \propto R^{-4}$. Finally, for intermediate values of $R \gtrsim 2$, on shell production of $A^\prime$ implies that the production rate is independent of $\alpha_D$, provided that the dark photon dominantly decays invisibly. In this case, smaller values of $\alpha_D$ increase the $y$-sensitivity of accelerator probes, while the thermal target remains unchanged. In this sense, larger values of $\alpha_D$ are a conservative choice for highlighting the coverage ability of low-energy probes to much of the cosmologically-motivated regions of parameter space. 


\temporarypagebreak
\subsection{Thermal Dark Matter Through other New Vector Interactions}


\begin{figure}[h!]
\begin{center}
\begin{subfigure}[t]{0.48\hsize}
\includegraphics[width=\hsize]{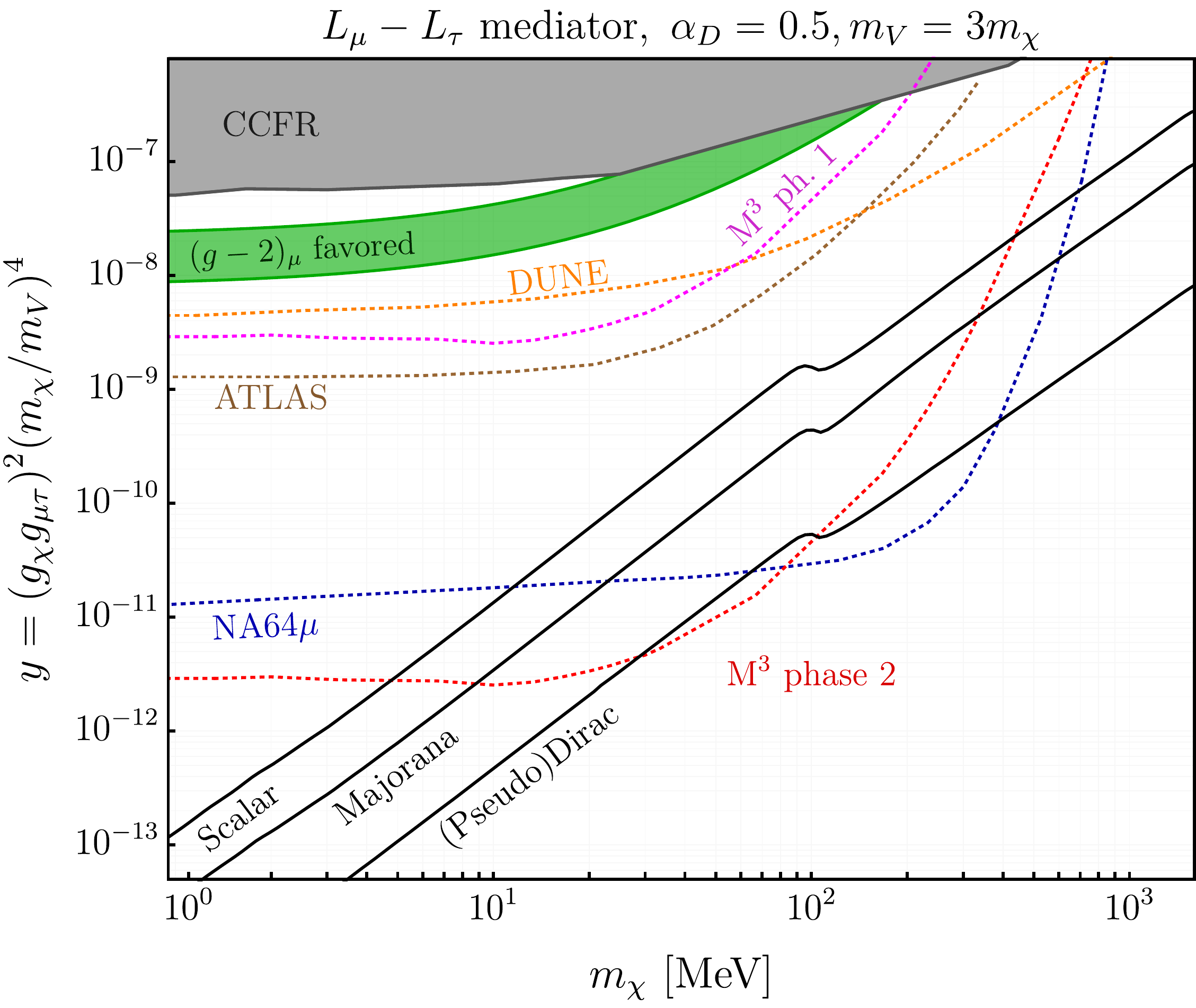}
\caption{$U(1)_{L_\mu-L_\tau}$, $\alpha_D=0.5$ 
}
\label{sfig:MuminusTau}
\end{subfigure}
\hfill
\begin{subfigure}[t]{0.49\hsize}
\includegraphics[width=\hsize]{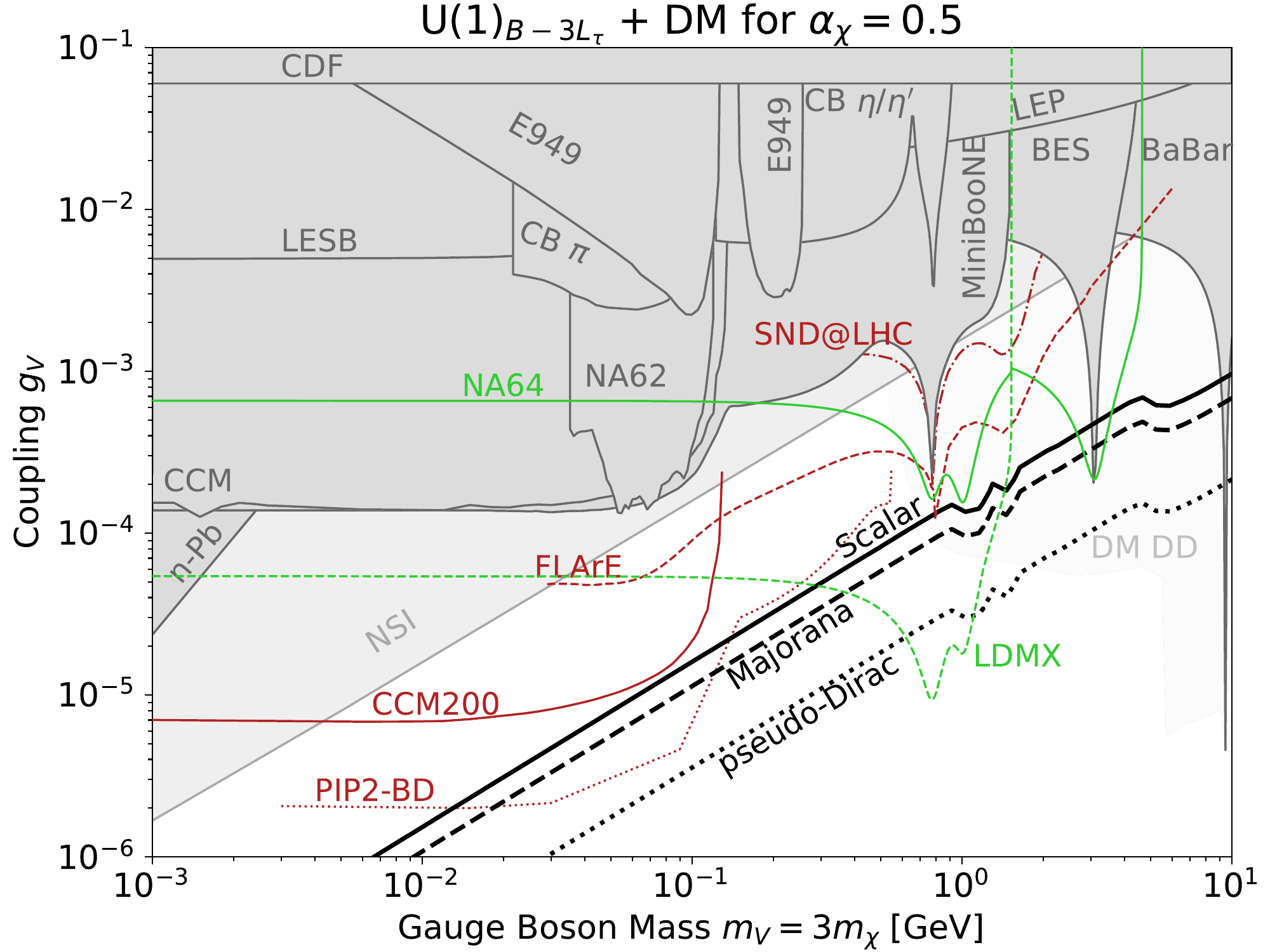}
\caption{$U(1)_{B-3L_\tau}$, $\alpha_D=0.5$
}
\label{sfig:Bminus3LTau}
\end{subfigure}

\begin{subfigure}[t]{0.49\hsize}
\includegraphics[width=\hsize]{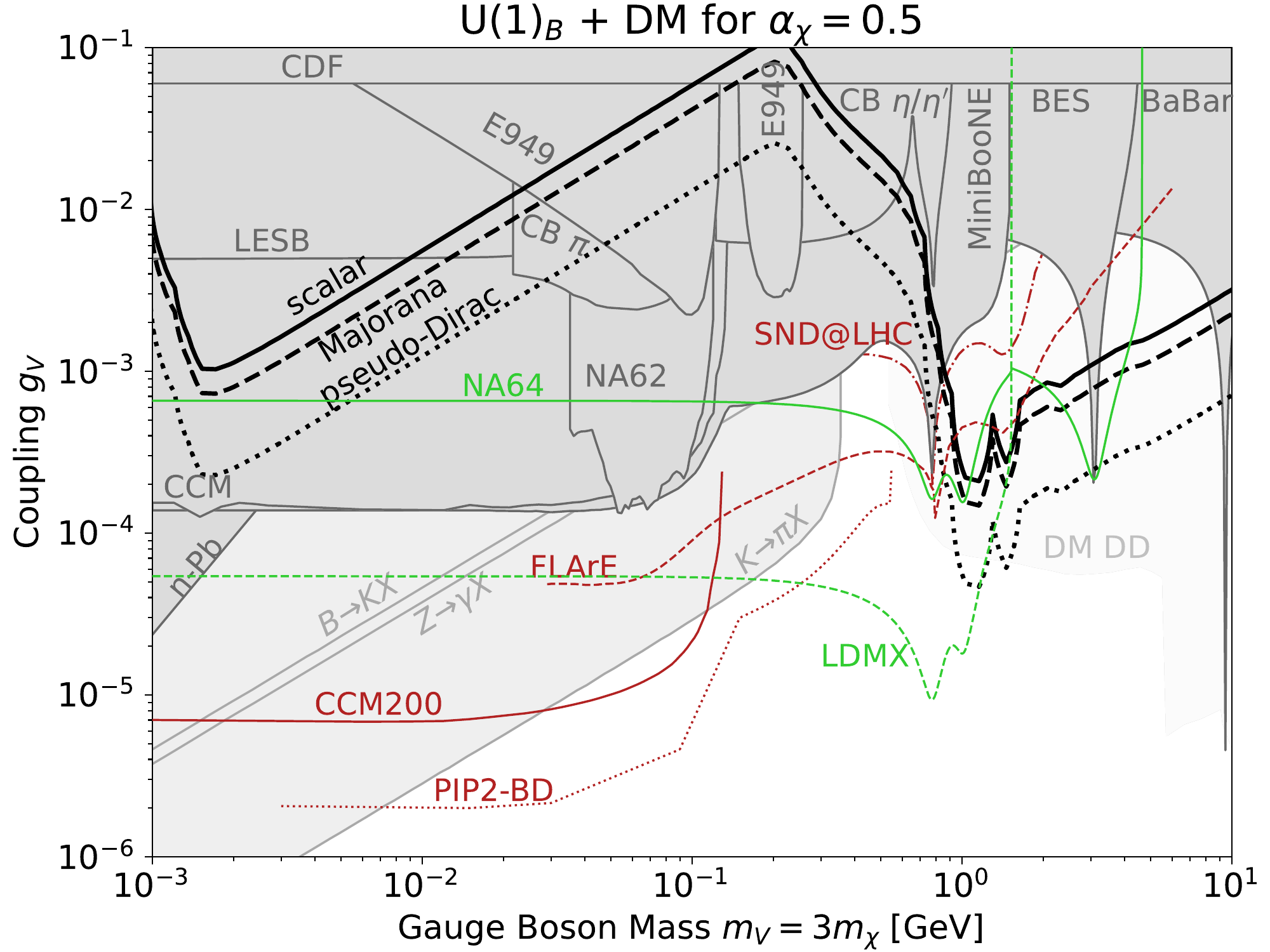}
\caption{$U(1)_{B}^{(*)}$, $\alpha_D=0.5$ 
}
\label{sfig:BaryonTargets_LargeaD}
\end{subfigure}
\hfill
\begin{subfigure}[t]{0.49\hsize}
\includegraphics[width=\hsize]{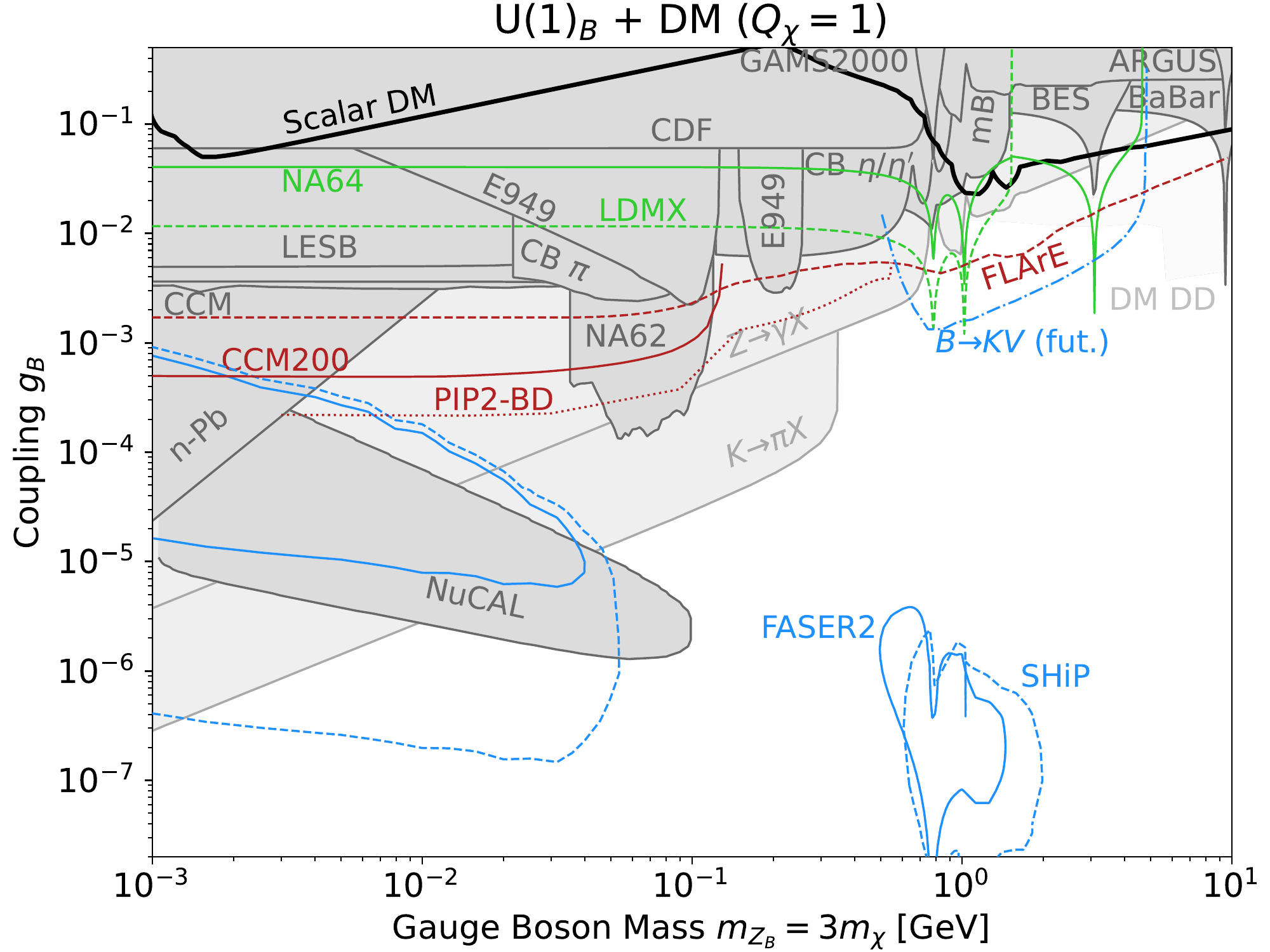}
\caption{$U(1)_{B}^{(*)}$, $\alpha_D=\alpha_B$  
}
\label{sfig:BaryonTargets_SmallaD}
\end{subfigure}
\end{center}
\caption{Thermal targets for several examples of a mediator $V$ coupled to SM global symmetries.  In the top four panels, we choose the conventional benchmark mass ratio $R=m_{V}/m_{\chi}=3$ and $\alpha_D=0.5$, and the $y$-axis is given by $y=\alpha_D \epsilon_V^2 R^-4$, where SM couplings are given by $\epsilon_V e Q$ for particles of charge $Q$ under the specified global symmetry. In panel \subref{sfig:BaryonTargets_SmallaD} we instead take $\alpha_D=\alpha_B$, as motivated if the coupling of the new gauge boson is intrinsically small (and comparable) for both dark matter and SM quarks.  The symmetries considered in \subref{sfig:MuminusTau}-\subref{sfig:Bminus3LTau} are anomaly-free, while the baryon number coupling in \subref{sfig:BaryonTargets_LargeaD} and \subref{sfig:BaryonTargets_SmallaD} is anomalous and therefore subject to additional high-energy constraints \cite{Dror:2017ehi}.
\label{fig:globalVectorsThermal}}
\end{figure}



In Figs~\ref{sfig:Bminus3LTau} and \ref{sfig:BaryonTargets_LargeaD} we present the projected exclusion bounds for the search for hadrophilic DM scattering signatures in the CCM200~\cite{CCM:2021leg,Aguilar-Arevalo:2021sbh}, FLArE~\cite{Batell:2021snh}, PIP2 BD~\cite{Toups:2022yxs}, and SND@LHC~\cite{Boyarsky:2021moj} experiments, as well as the relevant future bounds from missing energy/momentum searches at NA64~\cite{Gninenko:2016kpg,Schuster:2021mlr} and LDMX~\cite{Akesson:2022vza}. These have been obtained for the $U(1)_{B-3\,L_\tau}$ (left panel) and $U(1)_B$ (right) gauge boson $V$ coupled to the DM particles with $\alpha_D = 0.5$. In the plots, we assume that $m_V = 3\,m_\chi$. We also present the relic target lines for scalar, Majorana fermion, and pseudo-Dirac fermion DM particles. In the $U(1)_B$ case, we take into account suppressed annihilations into electron-positron pairs via loop-induced kinetic mixing, $\epsilon = e g_V/16\,\pi^2$. Instead, for the $U(1)_{B-3\,L_\tau}$ model, light DM species can efficiently annihilate into SM neutrinos. In the plots, past bounds are shown with a gray-shaded region following Ref.~\cite{Batell:2021snh}. These include anomaly-driven bounds from rare kaon, $B$ meson, and $Z$ decays shown in light gray for the $U(1)_B$ case~\cite{Dror:2017ehi,Dror:2017nsg}. These can be more straightforwardly avoided in the $U(1)_{B-3\,L_\tau}$ scenario by introducing additional heavy right-handed neutrinos. In the latter case, however, further bounds arise from neutrino non-standard interactions (NSI)~\cite{Heeck:2018nzc,Han:2019zkz}. We also show current DM direct detection (DD) experiments which, however, apply only to the scalar DM scenario.

Non-minimal signals in the dark sector might naturally arise when the new physics species decay both visibly and invisibly into DM particles with similar branching fractions. We illustrate this in Fig.~\ref{sfig:BaryonTargets_SmallaD} for the hadrophilic vector portal in which the DM sector is coupled to the SM via a $U(1)_B$ gauge boson. Here, we assume that the complex scalar DM particle $\chi$ carries $Q_\chi=1$ charge with respect to the $U(1)_B$ group, similarly to the SM quarks. This results in a common value of the relevant coupling constant, $\alpha_D = \alpha_B$, and non-negligible branching fractions of the dark vector into both SM quarks and $\chi$s. In the plot, we also assume that $m_V = 3\,m_\chi$. Hence, the parameter space of the model can be constrained by both future searches for DM scattering events, as shown with the projected exclusion bounds for the CCM200, FLArE, and PIP2 BD experiments, by missing energy/momentum searches at NA64 and LDMX, and by searches for visible decay products of $V$. The latter is illustrated with the expected future sensitivity line of $B\to KV$ searches~\cite{Dror:2017ehi}, and with the future bounds of the FASER 2 and SHiP experiments targeting highly displaced decays of $V$. We present the past bounds on the model with gray-shaded regions following Ref.~\cite{Batell:2021snh}.

\temporarypagebreak

\begin{figure}[ht]
\label{fig:higgs-mixed}
\includegraphics[width=0.49\hsize]{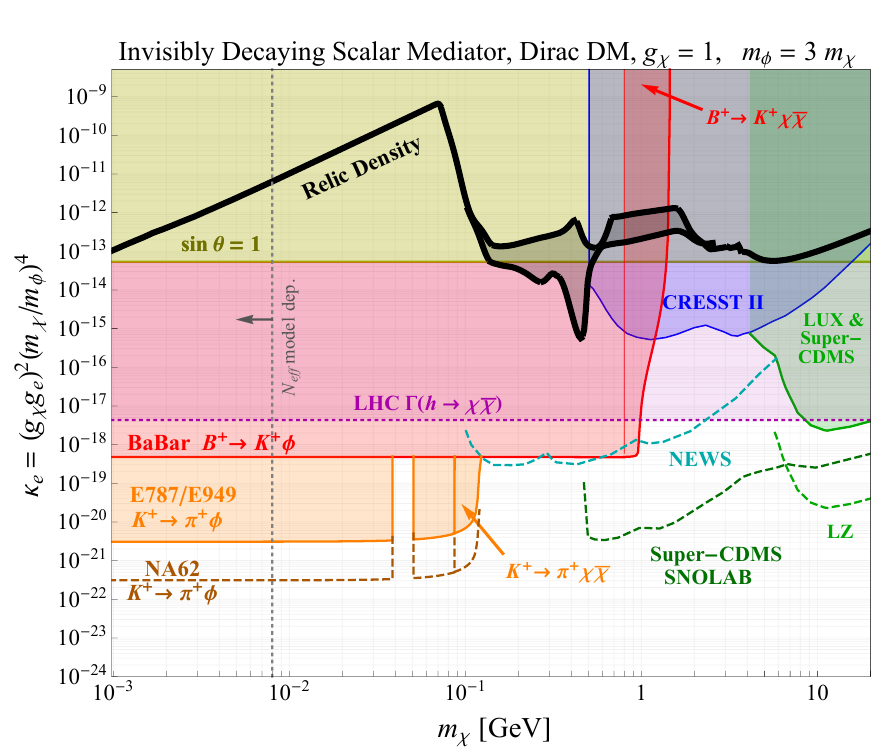}
\caption{Direct annihilation to SM particles through a light Higgs-mixed scalar \cite{Krnjaic:2015mbs}
\label{fig:hiigs_mixed_scalar_mediator}}
\end{figure}

\subsection{Scalar Mediators} 
\begin{figure}[h!]
\begin{center}
\includegraphics[width=0.45\hsize]{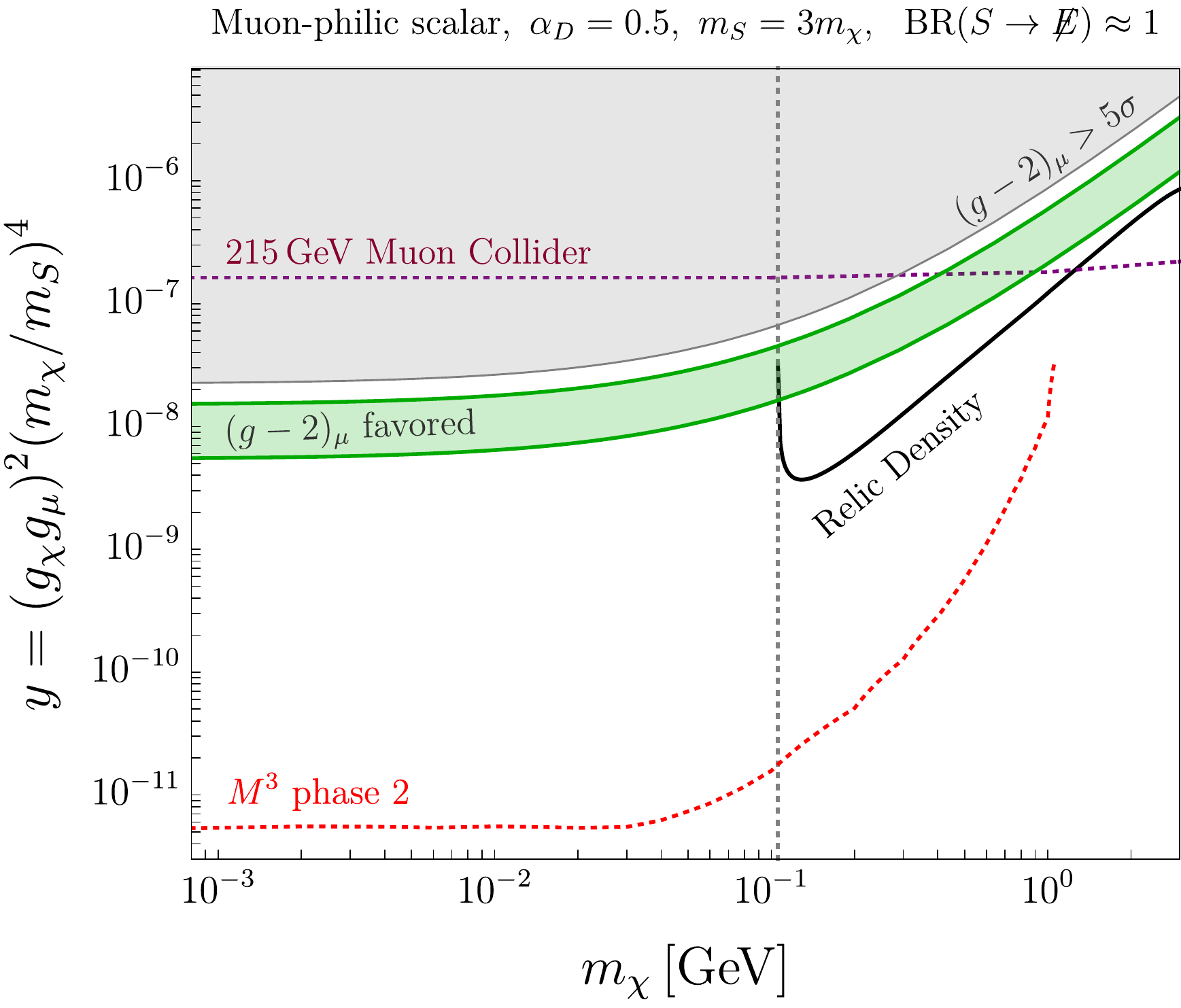}
\caption{Flavor Specific Scalar mediators couplings to muons. 
This model is one of only two viable 
scenarios for resolving the $(g-2)_\mu$ anomaly
via SM singlet particles below the
GeV scale \cite{Capdevilla:2021kcf}; the
other option is gauged $L_\mu - L_\tau$ shown in Fig. \ref{fig:globalVectorsThermal}.
}
\label{fig:muphilicscalar}
\end{center}
\end{figure}

Since SM fermion bilinears require an insertion of electroweak symmetry breaking, any Yukawa-like
coupling of a (pseudo) scalar mediator must either mix with the SM Higgs boson or 
arise from integrating out new electroweak states (e.g. a vectorlike fourth generation)  \cite{Batell:2009di}:

\begin{itemize}
    \item {\bf Higgs Mixing}: Mediators that mix with the Higgs acquire mass-porportional couplings to SM fermions.
    For light mediators, there are stringent bounds on the mixing parameter from various laboratory searches (particularly involving rare meson decays). Consequently,
    for any choice of light DM candidate that also couples to this mediator, there is no viable, unitary choice
    of couplings that can realize a thermal-sized cross section
    for predictive $s$-channel annihilation $\chi \bar \chi \to \bar ff$. This situation is summarized in Fig. \ref{fig:hiigs_mixed_scalar_mediator}.
    
    \item {\bf Flavor specific mediators:} A new spin-0 mediator can acquire flavor specific yukawa couplings 
    to SM fermions. Such interactions can arise by coupling the mediator to heavy new vectorlike states that mass-mix with SM fermions  \cite{Batell:2009di} or by mass-mixing the mediator with the states of 
    additional scalar doublets \cite{Egana_Ugrinovic_2019,Egana-Ugrinovic:2019wzj}. Depending on the flavor structure
    of the UV theory, this procedure can result in an arbitrary flavor pattern at low energies. In Fig. \ref{fig:muphilicscalar} we show a representative scenarion
    in which a muon philic scalar also couples to dark matter.

\end{itemize}

\temporarypagebreak
\subsection{Neutrino-portal DM production}

Given the paucity of information that we have about neutrinos and their properties, many extensions to the SM exist which posit new interactions with neutrinos, and, potentially, dark matter. The dimension-four neutrino portal, where the SM Lagrangian is augmented with a gauge-singlet fermion $N$ and an interaction term
\begin{equation}
{\cal L}\supset \lambda (LH)N+{\rm h.c.},
\end{equation}
allows for such interactions.
The sterile neutrino $N$ can be itself the DM candidate, or it can play the role of a meadiator and facilitate further interactions between the SM and the dark sector. The DM in any of these cases can be either secluded~\cite{Pospelov:2007mp,Tang:2016sib,Batell:2017rol,Escudero:2016ksa,Allahverdi:2016fvl,Campos:2017odj,Folgado:2018qlv}, annihilating into other dark states, or it may annihilate directly to SM particles~\cite{Macias:2015cna,Gonzalez-Macias:2016vxy,Batell:2017cmf,Berlin:2018ztp,Escudero:2016tzx,Escudero:2018fwn,Blennow:2019fhy}.
While standard freeze-out production is easily achieved if $N$ has sizeable interactions with the SM, freeze-in production has also been considered~\cite{Kang:2010ha,Abada:2014zra,Biswas:2016bfo,Chianese:2018dsz,Becker:2018rve,Bian:2018mkl,Chianese:2019epo,DeGouvea:2019wpf,Kelly:2020pcy,Kelly:2020aks}.

\textbf{Neutrino portal: $t$-channel:}
A minimal scenario where $N$ facilitates the interaction of a DM particle with the SM via $t$-channel annihilation was explored in Ref.~\cite{Batell:2017cmf}. The particle content is given by $N$, a Dirac fermion, that, with a scalar mediator $\phi$, couples to another Dirac field $\chi$ which constitutes the observed relic abundance of DM in the universe today. The interaction is given by
\begin{equation}
{\cal L}\supset y \phi\chi N+{\rm h.c.},
\end{equation}
The correct relic abundance can be achieved for reasonable couplings and for GeV-scale DM, allowing for searches at current and next-generation precision facilities. The coupling of the dark sector to the SM is controlled by the mixing angle between the active neutrinos and $N$,
\begin{equation}
U\sim \frac{\lambda v}{m_N}
\end{equation}
where $m_N$ is $N$'s (Dirac) mass.\footnote{Choosing $N$ to be Dirac decouples the light neutrino masses from $U$ and $m_N$, essentially allowing $U$ to be a free parameter. Addressing the origin of the light neutrino masses can be done in a way that does not alter this scenario~\cite{Batell:2017cmf,Bertoni:2014mva}.} This setup leads to the familiar case of a fourth neutrino mass eigenstate with $m_4\simeq m_N$ that is mostly $N$. The constraints on this scenario are less strong when $m_4>m_\chi$ so that the heavy neutrino decays invisibly. In addition, the dark matter in this case annihilates to light, mostly active neutrinos, with cross section
\begin{equation}
\sigma v\propto y^2\left|U\right|^2.
\end{equation}

\begin{figure}[htbp!]
\begin{center}
\includegraphics[width=0.5\columnwidth]{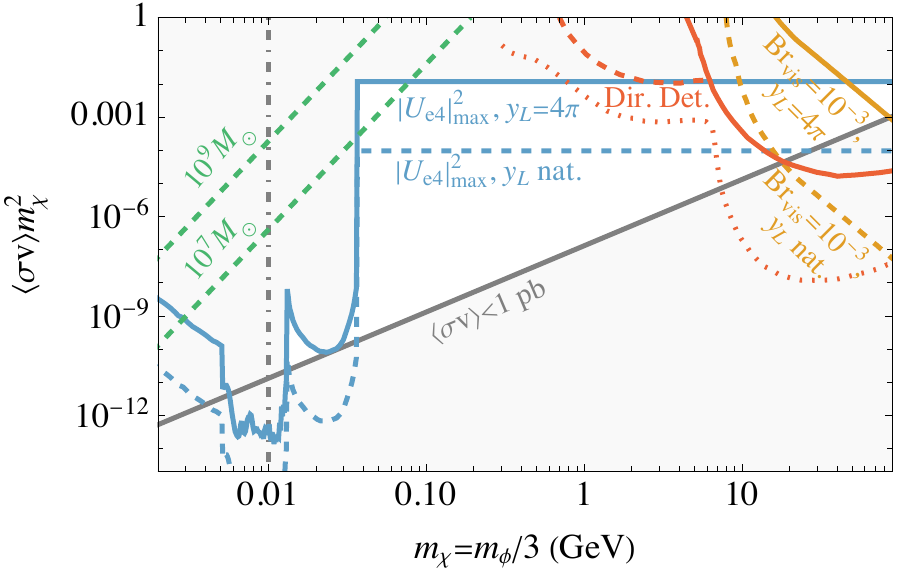}~\includegraphics[width=0.5\columnwidth]{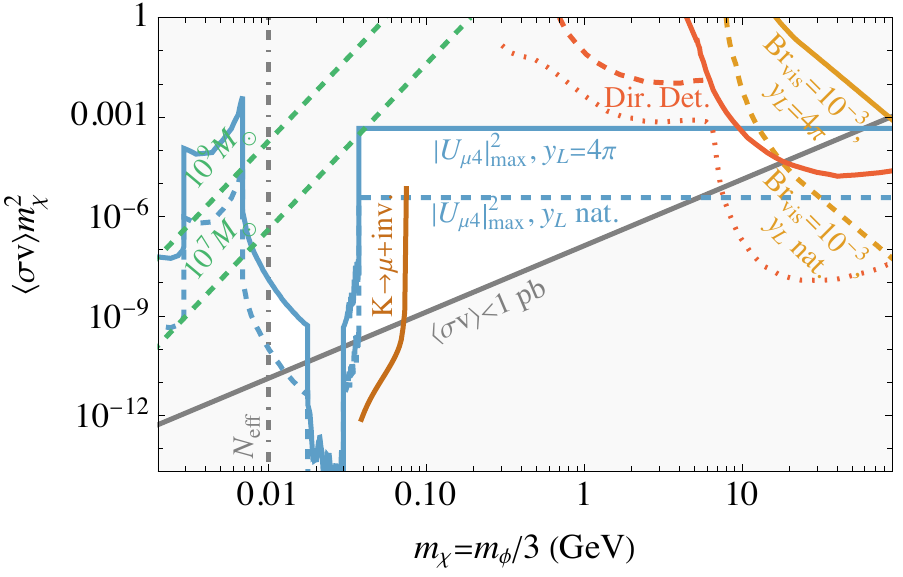}\\
\includegraphics[width=0.5\columnwidth]{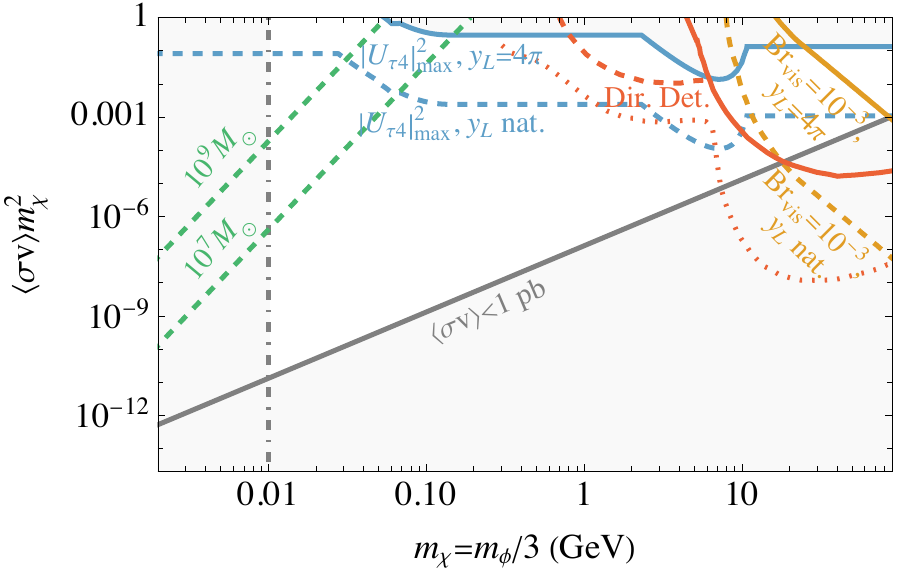}
\caption{The limits on the coupling between the sterile neutrino and the SM, parameterized by the dark matter anihilation cross section, in the $t$-channel neutrino portal scenario of Ref.~\cite{Batell:2017cmf}, assuming that the sterile neutrino $N$ dominantly mixes with $\nu_e$, $\nu_\mu$, or $\nu_\tau$. The masses of the new states are set to $m_\chi=m_\phi/3=m_4/10$. Constraints come from rare meson and lepton decays, $Z$ decays, dark matter direct detection, as well as small-scale dark matter structure. See~\cite{Batell:2017cmf} for details.}
\label{fig:neutrinoPortal_tchan}
\end{center}
\end{figure}

In Fig.~\ref{fig:neutrinoPortal_tchan}, we show a selection of constraints on this model in the $U$ vs. $m_\chi$ parameter space, fixing the sterile neutrino mass to be $3m_\chi$ and assuming that the sterile neutrino mixes dominantly with one flavor of SM active neutrino. Direct probes of this scenario are provided by rare decays of charged mesons and leptons that produce the sterile neutrino $N$ with a rate proportional to $\left| U\right|^2$. Once produced, $N$ generically decays invisibly into the dark sector, so that the characteristic signature of this scenario is charged meson or lepton decays where the momentum carried by the visible decay products differs from what is expected in the standard model. For instance, Ref.~\cite{Batell:2017cmf}, estimated the sensitivity of the E949~\cite{E949:2014gsn} experiment at BNL to the decay $K^+\to\mu^+\phi\chi$. NA62 will collect a much larger sample of charged kaon decays~\cite{NA62:2017ynf} and could cut into well-motivated parts of parameter space. In addition, the indirect detection signal from $\chi\bar\chi\to\nu\bar\nu$ offers another probe of this scenario~\cite{McKeen:2018pbb}.

\textbf{Neutrino portal: $s$-channel:} The right-handed neutrino $N$ can also serve as a mediator between the dark and visible sectors in $s$-channel annihilation.
In Refs~\cite{Escudero:2018fwn,Blennow:2019fhy,Ballett:2019pyw}, the authors introduce a sterile state $N$ as well as dark fermions charged under a secluded $U(1)^\prime$ gauge symmetry. The breaking of this symmetry by a complex dark scalar $\Phi$ gives rise to mixing between the dark leptons and the sterile neutrino $N$, and the breaking of EW symmetry gives rise to the mixing of $N$ with light neutrinos. As a consequence, dark sector particles can interact with SM neutrinos via the mediator of the $U(1)^\prime$ gauge group, $X^\mu$, as well as the physical scalar $\phi$, referred to as the dark photon and dark scalar, respectively. The interactions of the mediators with the SM are, to a good approximation, exclusively with light neutrinos, rendering the dark sector particles difficult to detect experimentally.

In what follows, we discuss one of the simplest realization of the model. Consider two Dirac fermions, $\chi$ and $N$. Only $N_L$ and $\chi_R$ are charged under the dark gauge symmetry such that the model is free of anomalies. Consider also that Lepton number is conserved only up to small terms related to neutrino masses. In this case, there exists a resulting "dark parity" $Z_2$ symmetry under which only $\chi$ is odd~\cite{Ma:2015xla} and $N$ is even. Therefore, neutrinos and $\chi$ particles do not mix and $\chi$ remains stable. Schematically, the model is defined by
\begin{equation}
    \mathcal{L} \supset \overline{N_L} \slashed{D}_X N_L + \overline{\chi_R} \slashed{D}_X \chi_R  - y_\chi \overline{\chi_L}(\Phi\chi_R) - y_N (\overline{N_L} \Phi^\dagger) N_R - \lambda (\overline{L}H) N_R + \text{h.c.},
\end{equation}
where $D_X^\mu = \partial^\mu - i g_X X^\mu$.
A Majorana mass for $\chi_L$ could induce a small splitting between the two dark matter states, and if large enough, the DM candidate would be a Majorana particle. The Majorana mass of $N_R$ would break lepton number, and if small, can generate the correct neutrino masses. This can happen at tree or loop level, depending on the number of generations of $N_R$ and $N_L$~\cite{Ballett:2019cqp}. Variations of this model exist, including gauge symmetries with non-trivial charge assignments~\cite{Gehrlein:2019iwl}.

\begin{figure}[htbp!]
\begin{center}
\includegraphics[width=0.75\columnwidth]{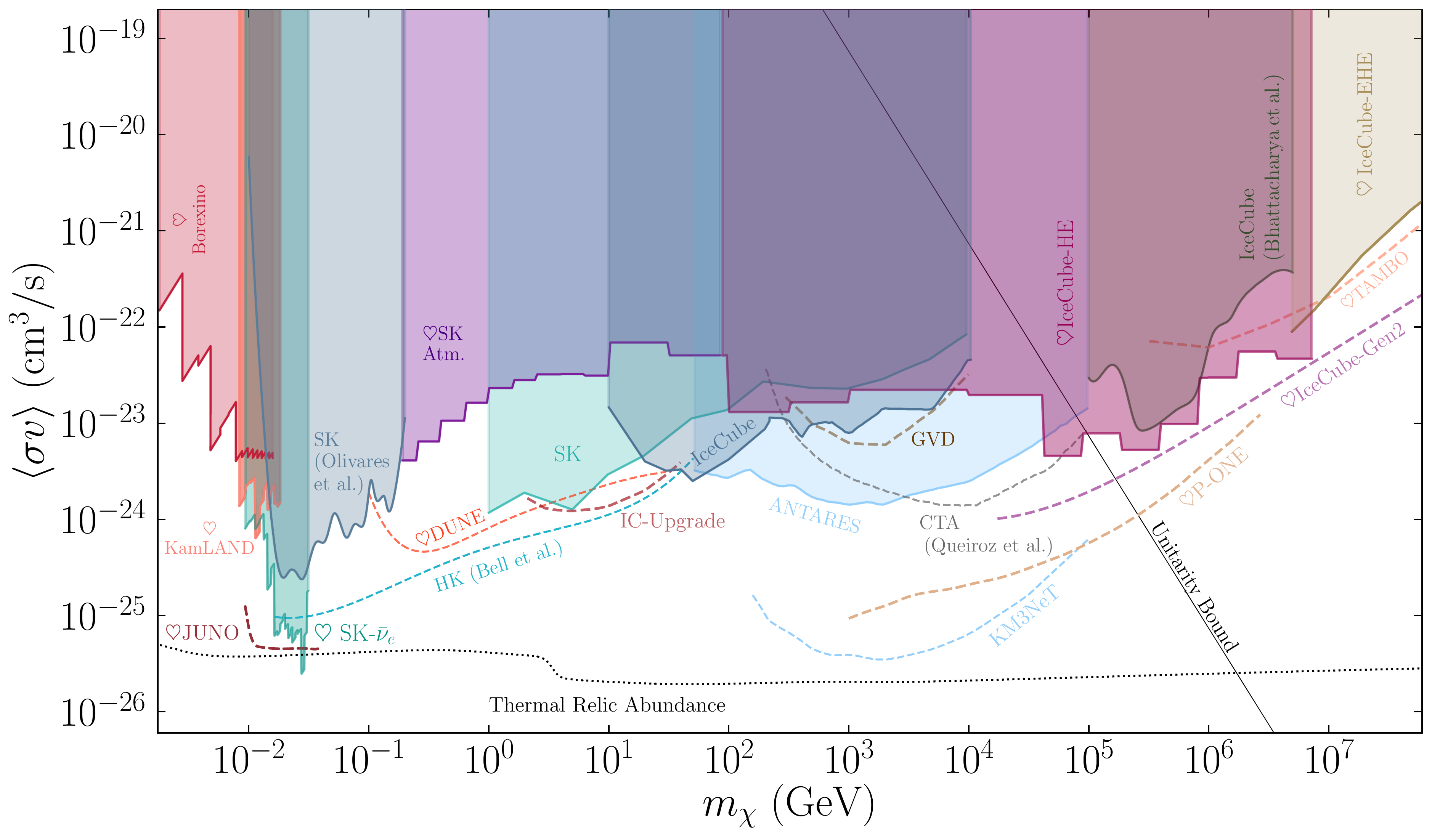}
\caption{The annihilation cross section for s-wave annihilation of fermionic dark matter into light neutrinos. The shaded regions are the existing experimental limits from searches of neutrino-lines at $E_\nu\simeq m_\chi$. Figure taken from \cite{Arguelles:2019ouk}.}
\label{fig:neutrinoPortal_schan}
\end{center}
\end{figure}

In this model, if $m_\chi > m_N, m_X, m_\phi$, DM particles annihilate into light neutrinos via either s-wave if $\chi$ remains a Dirac particle or p-wave if $\chi$ is split into two Majorana states. For s-wave, the annihilation cross section is approximately given by
\begin{equation}
\langle \sigma v\rangle_{\rm ann} \simeq  \frac{|U|^4 g^4_X}{8 \pi} \frac{m^2_{\chi}}{\left(4 m^2_{\chi}-m^2_{X}\right)^2}.
\end{equation}
Note that model-dependent limits apply depending on the size of active-heavy neutrino mixing, $U$, and the correct relic density can be achieved for $U \sim \mathcal{O}(10^{-3} - 10^{-2})$ for the largest allowed $g_X$. Several of the limits on $U$ are already shown in Fig.~\ref{fig:neutrinoPortal_tchan} and come from peak searches in meson decays, unitarity of the PMNS matrix, and leptonic decays. Due to the neutrino-philic interactions, $N$ is typically invisible in this model and therefore no beam-dump constraints apply.

In Fig.~\ref{fig:neutrinoPortal_schan}, we show model-independent limits on the annihilation cross section coming from indirect detection limints on $\overline{\chi}\chi \to \overline{\nu} \nu$~\cite{Arguelles:2019ouk}. We show limits on s-wave annihilation, noting that p-wave limits are worse. The experimental signature constitutes a monochromatic neutrino line from DM annihilation. Future projections for DUNE, Hyper-Kamiokande, JUNO, and future neutrino telescopes are also shown.

\textbf{Neutrino self-interactions:} If we allow for high-dimensional operators to couple to neutrinos, then further opportunities exist. For instance, augmenting the SM with a (potentially lepton-number-charged) gauge-singlet scalar $\phi$ and the interaction term $(LH)(LH)\phi$ at dimension-six provides new neutrinophilic self-interactions.  Connections to DM with such a scalar have been explored in a variety of contexts, where the relic abundance of the DM can either be set via a freeze-out~\cite{Kelly:2019wow} mechanism or freeze-in of sterile-neutrino DM through a modified Dodelson-Widrow~\cite{Dodelson:1993je} mechanism~\cite{DeGouvea:2019wpf,Kelly:2020pcy,Kelly:2020aks}.

Precision-frontier experiments, specifically those with intense neutrino fluxes, may search for the existence of a scalar $\phi$ via rare neutrino-scattering processes with large missing-transverse-momentum~\cite{Berryman:2018ogk,Kelly:2019wow,Kelly:2021mcd}.
\begin{figure}
    \centering
    \includegraphics[width=0.85\linewidth]{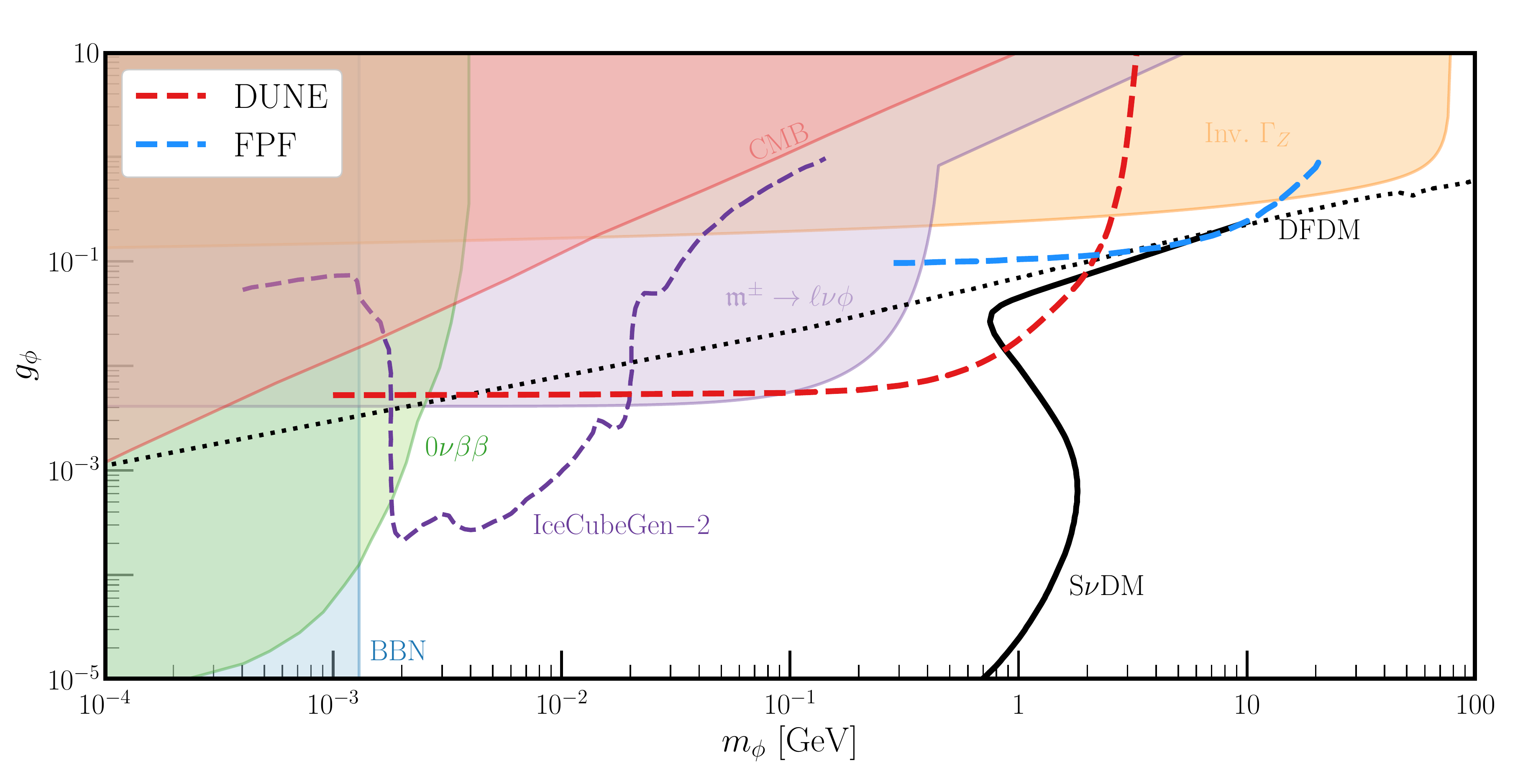}
    \caption{Summary of constraints and future sensitivity from precision-frontier experiments on neutrinophilic mediators $\phi$ and their connection to Dark Matter -- two DM target lines are shown, Dirac Fermion Dark Matter (dotted, black) and Sterile-neutrino Dark Matter (solid, black). \label{fig:NuphilicScalarDM}}
\end{figure}
Fig.~\ref{fig:NuphilicScalarDM} presents the prospects of several upcoming searches in this context, compared with existing constraints as a function of the scalar mass $m_\phi$ and its coupling to neutrinos $g_\phi$. Two benchmark DM targets are shown as black lines -- as a solid line, the sterile-neutrino freeze-in scenario for a ${\sim}7$ keV fermion DM; and as a dotted line, a Dirac-fermion DM that freezes out at its thermal relic abundance, with a mass $m_\chi \approx m_\phi/3$. Specifically, the upcoming DUNE and FPF experiments are capable of exploring this exciting GeV-scale mediator parameter space in the coming decade or so.

\temporarypagebreak

\subsection{Vector-Portal Model Variations with Non-Minimal Signals}\label{ssec:iDMandSIMPs}
As mentioned in Sec.~\ref{ssec:semivisible}, there are several classes of light dark matter models where \emph{unstable} particles in the dark sector play integral roles in the early-universe physics that dictates the DM abundance.  These DM excited states decay to a combination of  DM and SM final-states, so that detection of their visible products opens an avenue to exploring the nature and cosmological origin of DM.

\begin{figure}[htbp!]
\begin{center}
\begin{subfigure}[t]{0.49\hsize}
\includegraphics[width=\hsize]{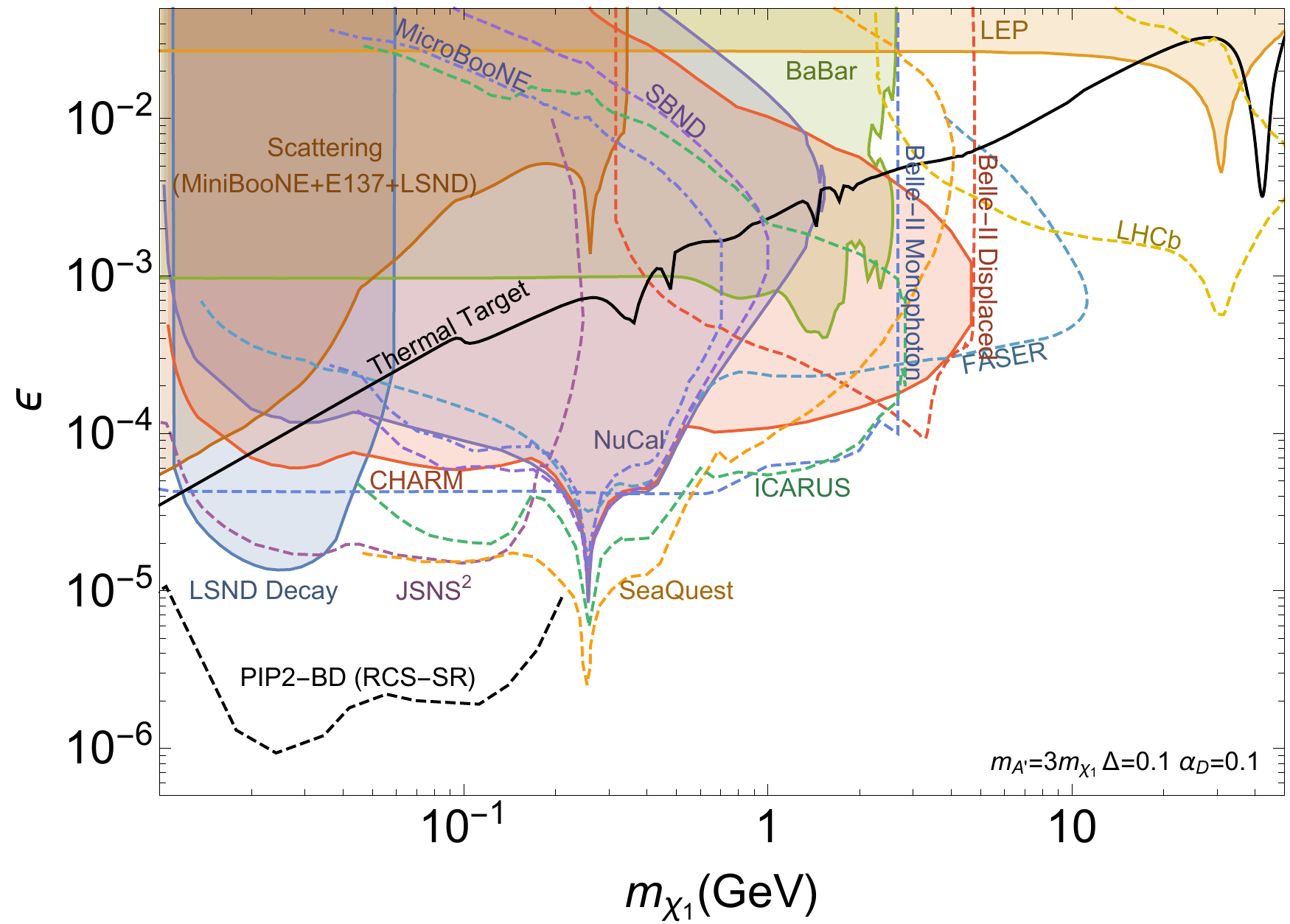}
\caption{Inelastic Dark Matter Example $\Delta=0.1$}
\label{sfig:iDM10}
\end{subfigure}
\hfill
\begin{subfigure}[t]{0.49\hsize}

\includegraphics[width=\hsize]{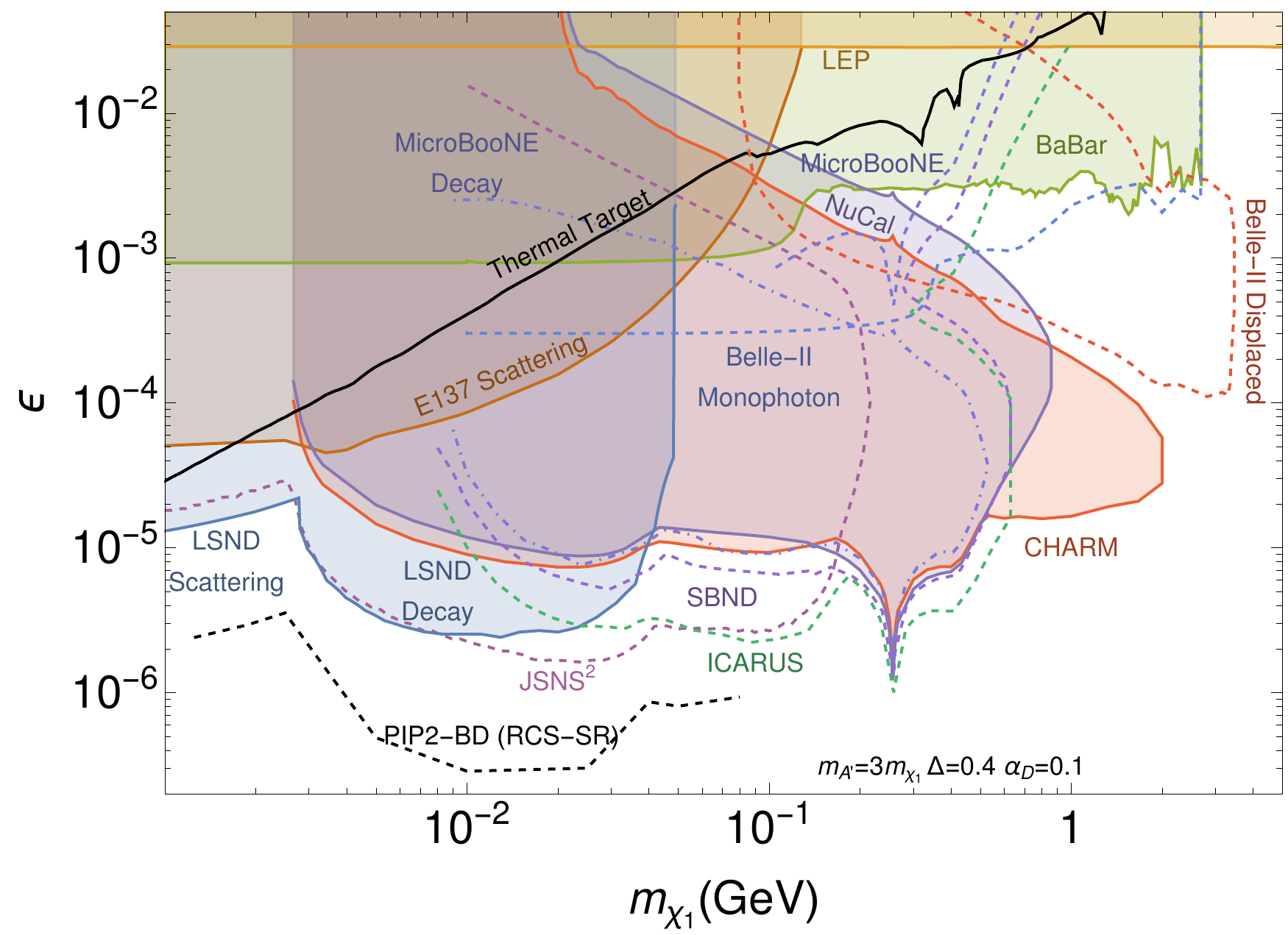}
\caption{Inelastic Dark Matter Example $\Delta=0.4$}
\label{sfig:iDM40}
\end{subfigure}

\begin{subfigure}[t]{0.49\hsize}
\includegraphics[width=\hsize]{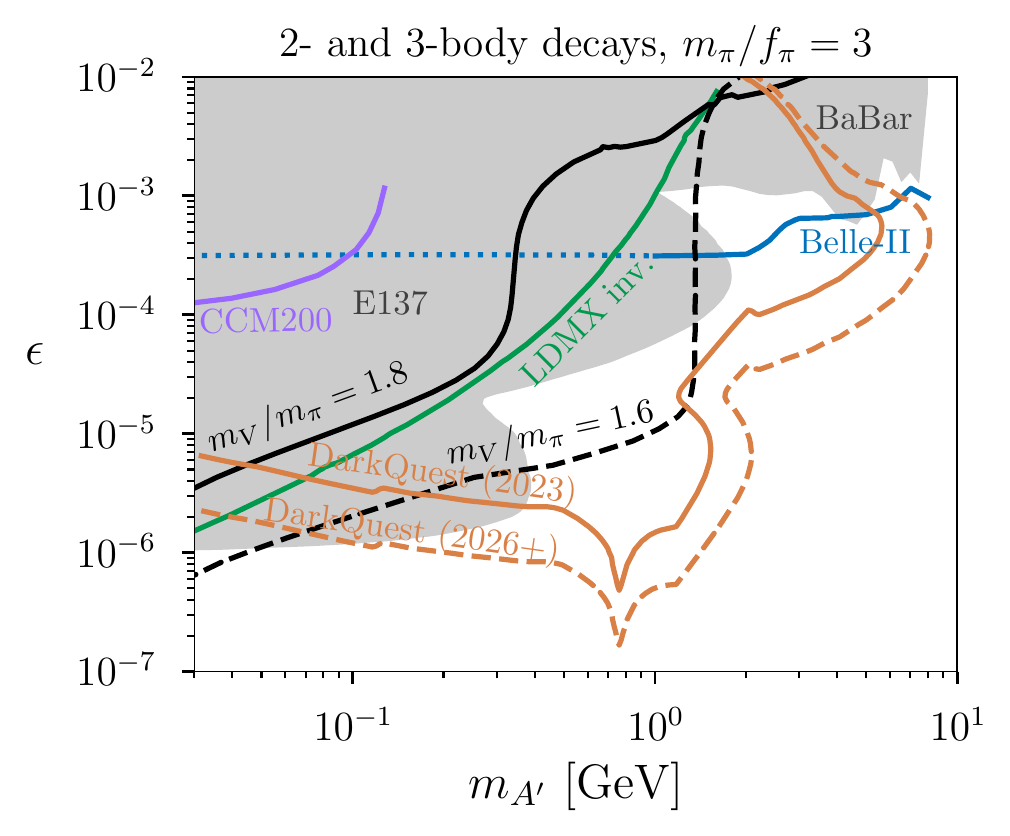}
\caption{SIMP Example}
\label{sfig:simp}
\end{subfigure}
\end{center}
\caption{Top Row: Inelastic DM with sufficient mass splitting to produce semi-visible signals.  Plots including DM thermal targets with $\Delta = 0.1$ (\subref{sfig:iDM10}) and 0.4(\subref{sfig:iDM40}) \cite{Izaguirre:2015zva, Izaguirre:2017bqb,Jordan:2018gcd,Berlin:2018jbm,Batell:2021ooj,Tsai:2019buq} 
Bottom row (\subref{sfig:simp}): Thermal targets and existing constraints in the Strongly Interacting massive particle (SIMP) model based on an $SU(3)$ dark sector gauge group coupled to the SM via the dark photon $A'$ with kinetic mixing $\epsilon$~\cite{Berlin:2018tvf}. The thermal targets are show for two ratios of dark vector meson to dark pion mass, $m_{V_D}/m_{\pi_D}$, and $\alpha_D=10^{-2}$, $m_{A'}/m_{\pi_D} = 3$ is taken throughout the plot. 
\label{fig:semiVisible}}
\end{figure}

Here, we discuss two simple benchmark models where the cosmological abundance of DM offers sharp predictions for experiment, and semi-visible searches offer dramatic sensitivity improvements relative to other kinds of DM searches: inelastic DM (iDM) and Strongly-Interacting Massive Particles (SIMPs). The former is a generalization of the vector portal models discussed previously, where the dark states consist of quasi-degenerate pair of particles $\chi_1$ and $\chi_2$, with the latter being slightly heavier, enabling its decays $\chi_2 \to \chi_1 + e^+ e^-$ ($\chi_1$ is the stable DM).
Figs.~\ref{sfig:iDM10} and \ref{sfig:iDM40} illustrate the sensitivity of scattering, kinematic, and semi-visible searches to iDM models with 10\% and 40\% mass splittings respectively between $\chi_2$ and $\chi_1$. 
These plots illustrate that past semi-visible searches place powerful constraints on the low-mass parameter regions of these models, and future analyses will place important constraints on iDM with few-GeV-scale masses.  

SIMPs, on the other hand, realize a qualitatively different thermal production mechanism which relies on $3\to 2$ number-changing reactions in the early universe to generate an adequate DM abundance~\cite{Hochberg:2014dra}. Such reactions (with a sufficiently large rate) are a feature of strongly-interacting gauge theories, akin to QCD~\cite{Hochberg:2014kqa}. Thus a dark sector with a confining gauge group realizes the SIMP mechanism; the DM in such a model is composed of analogs of pions, $\pi_D$; the analogs of vector mesons, $V_D$, constitute excited DM states that can decay visibly $V_D \to e^+ e^-$ or semi-visibly $V_D \to \pi_D e^+ e^-$~\cite{Berlin:2018bsc}. Sample thermal targets in the iDM and SIMP models are shown in Fig.~\ref{fig:semiVisible}.

\subsection{Millicharged particles}

\label{sec:millicharges}
Millicharged particles (mCP,  particles with small irrational electric charges) are motivated by the study of empirical charge quantization, grand unification theories, string theories, and dark sector models (especially the vector portal models) \cite{Dirac:1931kp,Pati:1973uk, Georgi:1974my, Shiu:2013wxa,Holdom:1985ag}. 
The millicharge can come from directly having a small hypercharge for the particles, or from a massless dark photon theory \cite{Holdom:1985ag}.
mCP is proposed to be a candidate of dark matter \cite{Brahm:1989jh, Feng:2009mn, Cline:2012is}, and could have significant consequences in cosmological measurements and help explain the EDGES anomaly \cite{Bowman:2018yin, Barkana:2018lgd, Munoz:2018pzp, Berlin:2018sjs, Slatyer:2018aqg, Liu:2019knx}.

\begin{figure}[h!]
\begin{center}
\includegraphics[width=0.95\hsize]{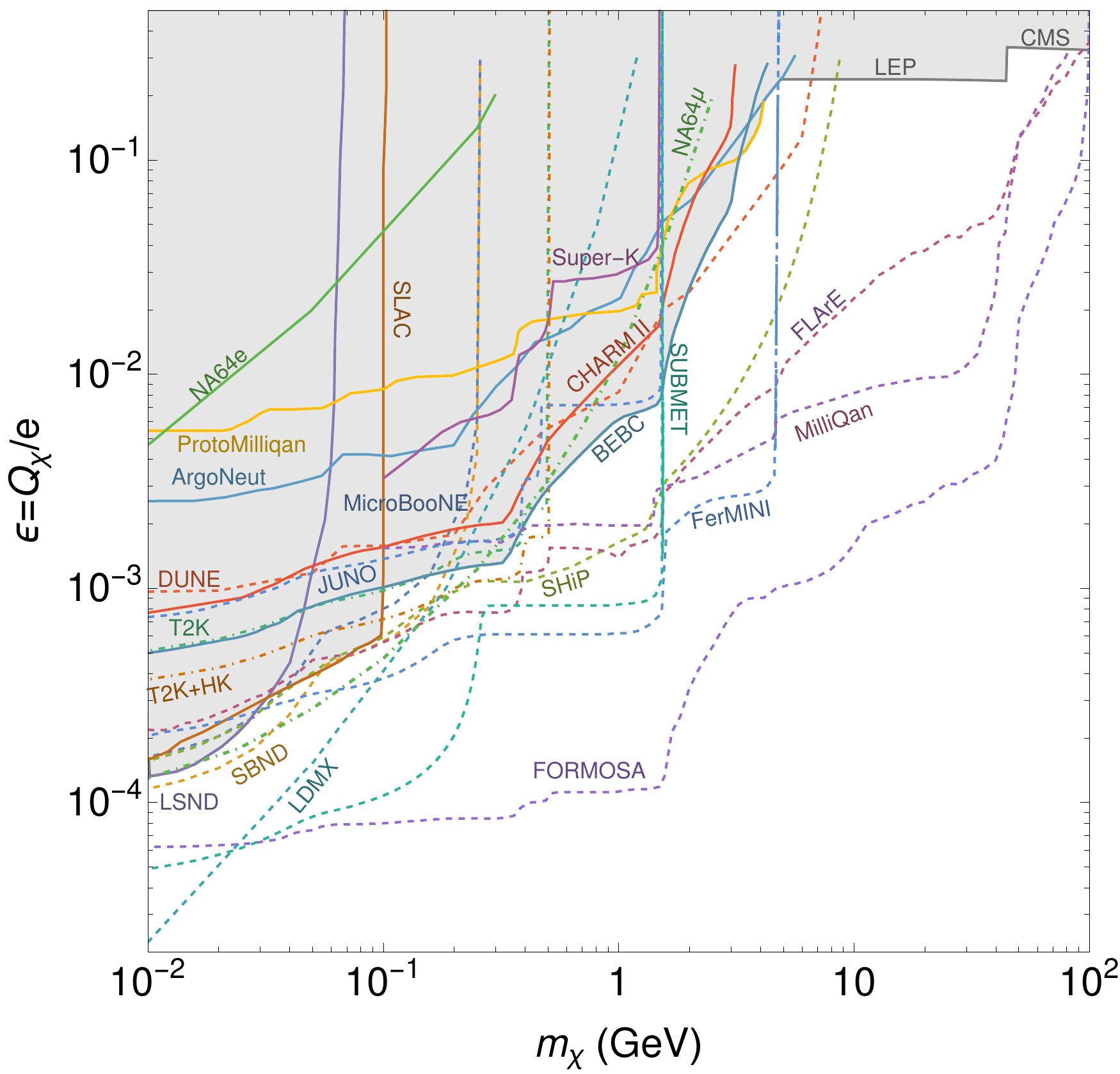}
\caption{
Constraints on mCP from previous searches include SLAC~\cite{Prinz:1998ua}, LEP~\cite{Davidson:2000hf, Badertscher:2006fm}, CMS~\cite{CMS:2012xi, Jaeckel:2012yz}, LSND~\cite{Magill:2018tbb}, ArgoNeuT~\cite{ArgoNeuT:2019ckq}, BEBC~\cite{Marocco:2020dqu}, Super-K limit on the diffuse supernova neutrino background~\cite{Plestid:2020kdm}, and the recent search by milliQan~\cite{Ball:2020dnx} are plotted in gray. Projections for milliQan at the HL-LHC~\cite{Haas:2014dda}, FerMINI~\cite{Kelly:2018brz}, SUBMET~\cite{Choi:2020mbk}, FORMOSA~\cite{Foroughi-Abari:2020qar}, and FLArE~\cite{Kling:2022ykt} are indicated as dashed curves. 
}
\label{fig:millicharge}
\end{center}
\end{figure}

One of the distinctive feature of mCP is the enhanced scattering cross-section in low-energy exchange. Scattering, sintillation, or missing energy/momentum searches are often the leading strategies for mCP studies.
Many accelerator experiments in RF6 can probe mCP. These searches include colliders \cite{Davidson:2000hf,CMS:2012xi,Jaeckel:2012yz,Ball:2020dnx,Liu:2018jdi,Feng:2022inv,Kling:2022ykt}, proton fixed-target and neutrino experiments \cite{Magill:2018tbb,Harnik:2019zee,ArgoNeuT:2019ckq,Marocco:2020dqu}, lepton fixed-target experiments \cite{Prinz:1998ua,Berlin:2018bsc,Gninenko:2018ter}, and dedicated searches \cite{Haas:2014dda,Kelly:2018brz,Foroughi-Abari:2020qar}. 
Similar to the accelerator searches, one can also study mCP produced in cosmic-ray hitting atmosphere and detected by large observatories \cite{Plestid:2020kdm,ArguellesDelgado:2021lek}.
The dedicated experimental efforts using of dedicated scintillation detectors, lead by milliQan \cite{Haas:2014dda}, FerMINI \cite{Kelly:2018brz}, SUBMET \cite{Choi:2020mbk}, and FORMOSA \cite{Foroughi-Abari:2020qar}, can search for millicharged particles down to $10^{-4}$ electric charges.
We include all of these studies in Fig.~\ref{fig:millicharge}.

\section{The Road Ahead}
In the past decade, a key goal of the light DM search effort has been broadly exploring the MeV to GeV dark matter parameter space compatible with simple thermal freeze-out mechanisms. The simplest, and most WIMP-like, mechanism for thermal freeze-out is annihilation to SM particles via an $s$-channel mediator that mixes with the photon (the dark photon portal mentioned above) or Higgs (excluded by observational constraints).  Thermal freeze-out through the dark photon portal has therefore emerged as a key benchmark model.  Because DM production at (semi)-relativistic kinematics drives both the dynamics of freeze-out and DM production at accelerators, the range of freeze-out interaction strengths (often parametrized by a dimensionless parameter $y$ related to the effective Fermi scale of the interaction) compatible with this mechanism is narrow, spanning a factor of $\sim 30$ at a given DM mass (black diagonal lines in Figure \ref{sfig:darkPhotonThermalR3}. 

\textbf{The Dark Matter New Initiatives (DMNI) BRN workshop highlighted, as a priority research direction achievable by the accelerator-based program, exploring interaction strengths singled out by thermal dark matter across the electron-to-proton mass range\cite{BRN}}.  OHEP DMNI funding has supported two experiments using different approaches to attain this goal, which requires a 10 to 1000-fold improvement in sensitivity over current searches:  In the near term, CCM200, a LAr-based re-scattering experiment at Los Alamos' LANSCE proton beam, will improve on present experiments' sensitivity by a factor of $\sim 10$, probing thermal DM to the scalar thermal milestone, for 10--100 MeV DM masses.  LDMX, a missing momentum experiment at SLAC's LESA electron beamline, is anticipated to begin running in the middle of the decade and will achieve the 1000-fold sensitivity improvements required to comprehensively explore the thermally motivated range of interaction strengths, including part of the resonant-annihilation region, for DM masses below $\sim 1/2$ GeV.  These small-scale, fixed target experiments are complemented by Belle II, where mono-photon analyses will explore the thermal milestones for DM masses between from 1/2 GeV to several GeV masses.  These experiments' sensitivity projections are illustrated in Fig.~\ref{sfig:darkPhotonThermalR3}.

This sensitivity is robust to many important model uncertainties, such as varying dark-sector couplings and the DM to dark-photon mass ratio (excepting a fine-tuned resonance-enhanced region), as illustrated in Fig.~\ref{sfig:darkPhotonThermalVaryR}. 

Fig.~\ref{fig:execDarkPhoton} also highlights the DM-search capabilities of many other experimental concepts outside the scope presently funded by DMNI.  Most of these leverage existing HEP investments in both accelerators and detectors, in the form of either analyses at multi-purpose detectors or modest improvements to existing beamlines. The breadth of the program is important for several reasons.  In particular: 
\begin{itemize}
\item \textbf{Multiple techniques with qualitatively different experimental backgrounds} are needed to assure a robust program; in the case of a positive signal, the complementary parameter-scalings and measurement capabilities of different approaches will allow a fuller characterization of the dark sector, including measurements of dark sector couplings, masses, and mediator interaction strengths with different SM particles. 
\item Beyond the dark photon portal, thermal-motivated milestones can also arise in more general models --- for example, those where a vector mediator couples weakly to a global symmetry of the SM or a scalar mediator couples to specific flavors.  Some of these, such as a vector coupled to $B-L$ (baryon number minus lepton number) give rise to qualitatively similar predictions to \ref{fig:execDarkPhoton} for all experiments. Alternatively, some models, including gauged $L_\mu - L_\tau$ or $B-3L_\tau$, feature mediators that do not couple appreciably to electrons, with DM annihilating into neutrinos, heavier leptons, or hadrons. These cases partially evade the sensitivity of electron-beam experiments in \ref{fig:execDarkPhoton}, and so experiments using \textbf{muon and proton beams} are critical to broadly testing these possibilities.  
\textbf{In particular, experiments at muon beams offer a particularly direct and comprehensive test of 
scenarios that resolve the muon g-2 anomaly with light new particles} \cite{Capdevilla:2021kcf}.  Some of these examples are highlighted in Figs.~\ref{fig:globalVectorsThermal} and \ref{fig:muphilicscalar}.  
\item In models where meta-stable particles in the dark sector (``DM excited states'') play important roles in DM cosmology, they can also be key to enabling discovery.  These models include inelastic DM with a large splitting and SIMP dark matter, and leveraging the semi-visible signals arising from decay of DM excited states can dramatically expand sensitivity to these models. For example, \textbf{semi-visible signals from excited state decay} are key to probing large-splitting inelastic DM at GeV-scale masses and the weak couplings (below the aforementioned thermal milestones) motivated by SIMP cosmology.   These sensitivities are highlighted in Fig.~\ref{fig:semiVisible}.
\item In addition to the bosonic portals emphasized above, there are also models where DM annihilates to the SM through the \textbf{neutrino portal}, which have qualitatively different experimental signals.  These are summarized in Figs.~\ref{fig:neutrinoPortal_tchan} and \ref{fig:neutrinoPortal_schan} .
\item Finally, a distinct avenue that has seen a recent surge of interest is the detection of \textbf{millicharged-particle production} both at fixed-target experiments and in LHC collisions.  Millicharged particles could be a small fraction of DM, while being too strongly coupled to be visible in direct detection.  Millicharged particles present a distinctive detector signature of very weak but continuous ionization, which can be exploited by dedicated detectors as well as searched for in accelerator-based neutrino detectors and missing energy/momentum searches. This parameter space and the new opportunities it motivates are summarized in Fig.~\ref{fig:millicharge}.
\end{itemize}

Each of the above-mentioned themes represents an exciting opportunity within light DM science.  
%

\subsection{Connections within RF6 and With Other Topical Groups and Frontiers}
\subsubsection{Experimental Synergy}
The breadth of the experimental program discussed in this whitepaper is considerable, with over 25 distinct future experiments reflected in the plots. Some of these are projections from dedicated experiments, while others are simply light-DM-focused analyses of data from experiments whose primary focus is in another topical group or frontier (e.g.~Belle II, COHERENT-STS, DUNE), or elsewhere in RF6 (e.g.~semi-visible searches in HPS or DarkQuest, experiments that mainly target minimal dark sector signals). In essentially all cases, the dedicated light DM experiments still benefit from cost-savings through use of existing infrastructure, accelerators, and/or hardware, and many have strong hardware synergies with other programs in HEP or with one another.  For example:
\begin{itemize}
    \item Many of the ``new experiment'' proposals nevertheless heavily leverage existing beamlines, beam dumps, and/or detectors.  For example the proposed SBN-BD uses the PIP-II linac and SBN near detector (SBND) at FNAL, while calling for construction of a new beam dump target station closer to the SBND; BDX requires a new detector facility but is operationally parasitic, as it would simply collect data while high-current experiments are operating in JLab's Hall A; LDMX will operate on the new LESA transfer line, which will be built at low cost as an offshoot of SLAC's LCLS-II XFEL. FLAre, FASER-II, FORMOSA, and MilliQAN would all operate as auxilliary detectors at the LHC. 
    Department of Energy investment in future beam dump
    experiments, including the DUNE near detector, will also offer new opportunities for probing light dark matter \cite{deRomeri:2020kno}

    \item Many of the experiments have technology synergies with flagship detectors in other frontiers. The LDMX ECal, for example, heavily leverages the design of the CMS HGCal and the LDMX analysis will also inform understanding of the HGCal detector's performance at the LHC.
    \item The FerMINI, FORMOSA, MilliQan, and SUBMET proposals all use essentially the same detector technology to search for millicharged particles, but probe different ranges of mass and charge by virtue of being situated at different locations relative to fixed-target and or collider interaction points.
        \item DarkQuest is an upgrade of the existing FNAL experiment SpinQuest to include a EM calorimeter (recycled from the PHENIX experiment), enabling broad sensitivity to semi-visible final states
\end{itemize}
In short, most of the opportunities presented here to search for light DM represent excellent value-added to ongoing programs, in return for investments that are modest on the scale of modern HEP. 

Theoretical physicists have been instrumental not only to the development of many of the dark matter models that motivate this research program, but also to the development of new experimental approaches and maximizing their capabilities (a Snowmass whitepaper \cite{Essig:2022yzw} further discusses some examples of the contributions of theorists in developing new experiments in light DM and related fields). Further support for theory research is particularly vital in this research area.  Over the next decade, two areas of theory particularly relevant to RF6 are the close collaboration between theorists and experimentalists, at all stages of experimental development from conceptual proposals through design optimization and science analysis, and the development of new models for dark matter and their connections to other problems in particle physics.  

\subsubsection{Complementarity to Other Probes of Dark Matter}
To close this whitepaper, we note several important points of complementarity with other experiments that will shed light on dark matter in the coming decade. We begin by discussing synergy with other topics in RF6, then move outward to accelerator-based Energy-Frontier DM searches at accelerators and finally to light DM searches in the Cosmic Frontier. 

\paragraph{Within RF6:}
As was noted above, the search for semi-visible signals of DM is a cross-cutting topic within the dark sector program. These searches address predictive models for DM that incorporate non-minimal structure in the dark sector (a focal area of \cite{BI3}) and several of the relevant experiments are in fact optimized or conceived to search for minimal dark-sector portals (reviewed in \cite{BI2}).  

A second important facet of complementarity within RF6 is that some DM models allow couplings to the SM well below our thermal milestones because the \textbf{leading interaction controlling freeze-out occurs within the dark sector, with mediator decays to SM particles playing an important role}.  Examples include DM annihilation into secluded scalars which decay into SM particles \cite{Krnjaic:2015mbs} and models such as hidden-sector Forbidden DM, Not-Forbidden DM, and Kinetically Decoupling DM \cite{DAgnolo:2015ujb,Cline:2017tka,Fitzpatrick:2020vba,Fitzpatrick:2021cij} in which the DM annihilates into dark photons with weaker couplings than we consider here.  In each of these cases, the DM itself may be quite challenging to see (by any means, including accelerators, direct detection, and indirect detection) but \textbf{much of the allowed parameter region can be explored by upcoming searches for the mediator}, which under the assumptions of these models should decays directly to SM final states via one of the minimal portals considered in \cite{BI2}.  In these scenarios, the first evidence of the dark sector and a powerful clue to the nature of DM will be revealed through searches for visibly decaying mediators.

\paragraph{Complementarity to Broader RF Physics Goals:}

Independently of their connection to dark matter, many of the mediators and dark sector states outlined in this document have overlap
with physics goals in the larger RF portfolio. Of particular interest
are muon-philic scalar and $L_\mu - L_\tau$ vector shown respectively in Figs. \ref{fig:muphilicscalar} and \ref{sfig:MuminusTau}, which are the only viable scenarios to resolve muon $g-2$ anomaly with light new particles.
\textbf{New muon beam missing-energy/momentum experiments offer the only experimental opportunity that comprehensively tests these key $g-2$ motivated milestones.}
\cite{Capdevilla:2021kcf}

\paragraph{Energy Frontier:}
There is a strong conceptual similarity between the dark photon models we have discussed and the simplified models of DM-SM interaction frequently analyzed in mono-jet searches at the LHC and mono-photon searches at future $e^+e^-$ colliders. The studies differ not only in the search methods, but in the mass scale of physics they are sensitive to, with high-energy colliders focusing primarily on DM interacting through \emph{heavier mediators} with \emph{SM-like or larger interaction strengths}, while our focus has been on \emph{lighter mediators} with \emph{interactions with SM matter parametrically weaker than those of SM forces}.  Each of these represents a natural set of assumptions for weak-scale or sub-GeV-scale dark matter, respectively.  (It is also the case that somewhat different ratios of SM interaction strengths are usually considered in the two communities, but this is a more incidental difference).  

It is worth calling attention to an interesting gap between these two mass regions: \textbf{mediator and/or DM masses in the few tens of GeV} are kinematically inaccessible to intensity-frontier experiments, and can be challenging to trigger on in multi-TeV hadron collisions. They are, nonetheless, viable, and indeed motivation for searching in this mass range actually preceded the recent interest in sub-GeV models (see e.g.~ \cite{Strassler:2006im,Feng:2008ya,Feng:2009mn}). From the perspective of this whitepaper, it is notable that \emph{the thermal milestones we have focused on extend to masses above a GeV, and are not yet explored for DM and mediator masses between the B-factory kinematic limit and the $Z$-pole}.  The challenge of exploring this parameter space likely requires creative new analyses of energy-frontier data, and would be an exciting complement to the program discuss here. 

\paragraph{Cosmic Frontier:}
Among cosmic-frontier probes of DM, low-threshold direct detection (see \cite{CF1_Essig:2022dfa}) is the most directly comparable to the searches discussed here.  Although both detection techniques rely on a coupling of the dark sector to familiar matter, and they explore overlapping ranges of models, the two approaches are complementary in several respects. First, they are sensitive to different properties of DM and the dark sector: Direct detection can only probe the DM itself via interactions that have elastic kinematics to within a part in $10^{-6}$; accelerator production explores the whole dark sector. Conversely, direct detection probes the DM's cosmological abundance and stability, while accelerators strictly speaking can only identify ``DM candidates'' without establishing their presence in the Galactic DM halo.  Moreover, each is well suited to measuring different DM properties, with the combination of the two being particularly powerful.
 
 Second, the two approaches operate at very different kinematics, with direct detection probing momentum transfers of ${\cal O}(10^{-3})$ times the DM mass and accelerators probing momentum transfers of ${\cal O}(\rm few)$ times the DM mass.  For thermal freeze-out models, the latter is generally advantageous --- for elastic scalar DM, the direct detection cross-section is velocity-independent, but for Majorana DM \ref{eq:majorana} it is suppressed by $v^2$,  while pseudo-Dirac (\ref{eq:pseudoDirac}) and inelastic scalar (\ref{eq:scalar}) DM have predominantly mass-off-diagonal interactions so that direct detection can only find these models via a one-loop diagram proportional to $y^2$ rather than $y$ \cite{Berlin:2018jbm}.  While elastic scalar thermal DM can likely be explored within the decade by direct detection, the other comparably motivated thermal scenarios are beyond reach.\footnote{However, there has 
 been progress developing conceptual ideas for improving direct detection sensitivity to the Majorana target in the long term \cite{Kahn:2020fef}}  (Likewise, models with ultra-light mediators have Coulombic scattering rates that are \emph{enhanced} at low velocities --- these can be visible in direct detection but not at accelerators, accentuating their complementarity of the two approaches). 
 
 If model-independent comparisons to direct detection are challenging, other cosmic probes of DM are even less comparable. However, they should be noted as complementary avenues to learning about similar DM sectors. Interesting complementary avenues include future precision measurements of the effective number of neutrino species $N_{\rm eff}$ \cite{Nollett:2013pwa,Planck:2018vyg}, which can be sensitive to light DM, and of small-scale structure which can constrain or shed light on DM self-interactions and to DM-baryon
 scattering, albeit for larger than thermal
  cross sections \cite{Bechtol:2019acd}.  In the coming decade, it will be especially exciting to see all of these probes of light DM come together, and the interpretation of a discovery in any one arena will be informed by a wealth of data in the others. 





\bibliographystyle{JHEP}
\bibliography{references}  

\end{document}